\documentclass[reprint,amsmath,amssymb,aps,prb,superscriptaddress]{revtex4-1}
\pdfoutput=1
\usepackage{array}
\usepackage{bm}
\usepackage{graphicx}
\usepackage{hyperref}
\hypersetup{colorlinks=true,allcolors=blue}
\usepackage{natbib}
\usepackage{newtxtext}
\usepackage{newtxmath}

\begin{document}

\title{Quantum spin liquid and magnetic order in a two-dimensional non-symmorphic lattice: \\ considering the distorted Kagome lattice of Volborthite}

\author{Li Ern Chern}
\affiliation{Department of Physics, University of Toronto, Toronto, Ontario M5S 1A7, Canada}

\author{Kyusung Hwang}
\affiliation{Physics Department, Ohio State University, 191 W Woodruff Avenue, Columbus, Ohio 43210, USA}

\author{Tomonari Mizoguchi}
\affiliation{Department of Physics, University of Tokyo, 7-3-1 Hongo, Bunkyo-ku, Tokyo, 113-0033, Japan}

\author{Yejin Huh}
\affiliation{Department of Physics, University of Toronto, Toronto, Ontario M5S 1A7, Canada}

\author{Yong Baek Kim}
\affiliation{Department of Physics, University of Toronto, Toronto, Ontario M5S 1A7, Canada}
\affiliation{Canadian Institute for Advanced Research/Quantum Materials Program, Toronto, Ontario M5G 1Z8, Canada}
\affiliation{School of Physics, Korea Institute for Advanced Study, Seoul 130-722, Korea}



\begin{abstract}
The Kagome-lattice-based material, Volborthite, $\mathrm{Cu_3 V_2 O_7 (OH)_2 \cdot 2 H_2 O}$, has been considered as a promising platform for discovery of unusual quantum ground states due to the frustrated nature of spin interactions. Here we explore possible quantum spin liquid and magnetically ordered phases in a two-dimensional non-symmorphic lattice described by $p2gg$ layer space group, which is consistent with the spatial anisotropy of the spin model derived from density functional theory (DFT) for Volborthite. Using the projective symmetry group (PSG) analysis and Schwinger boson mean field theory, we classify possible spin liquid phases with bosonic spinons and investigate magnetically ordered phases connected to such states. It is shown, in general, that only translationally invariant mean field states are allowed in two-dimensional non-symmorphic lattices, which simplifies the classification considerably. The mean field phase diagram of the DFT-derived spin model is studied and it is found that possible quantum spin liquid phases are connected to two types of magnetically ordered phases, a coplanar incommensurate $(q,0)$ spiral order as the ground state and a closely competing coplanar commensurate $(\pi,\pi)$ spin density wave order. In addition, periodicity enhancement of the two-spinon continuum, a signature of symmetry fractionalization, is found in the spin liquid phases connected to the $(\pi,\pi)$ spin density wave order. We discuss relevance of these results to recent and future experiments on Volborthite.
\end{abstract}


\maketitle

\tableofcontents

\section{\label{introduction}Introduction}

The interplay between lattice structure and spin exchange interactions is the defining characteristic of frustrated quantum magnets that hold the promise for exotic quantum ground states such as quantum spin liquid phases.\cite{nature08917} A prominent example of frustrated magnets is the Kagome lattice system with localized $S=1/2$ moments. There exist several materials that host $S=1/2$ local moments on various versions of the Kagome lattice. For example, a number of experiments\cite{PhysRevLett.98.077204,PhysRevLett.98.107204} observe signatures of a possible spin liquid ground state in the isotropic Kagome antiferromagnet,\cite{ja053891p} Herbertsmithite $\mathrm{Zn Cu_3 (OH)_6 Cl_2}$. Another material, Volborthite $\mathrm{Cu_3 V_2 O_7 (OH)_2 \cdot 2 H_2 O}$, has a distorted Kagome lattice \cite{JPSJ.70.3377,ncomms1875} and a magnetic order appears below 1K. Clearly, the difference in lattice structure plays an important role in the determination of the quantum ground state.

Volborthite, however, shows a rather complex response to an external magnetic field and its full magnetic phase diagram is still under investigation.\cite{JPSJ.78.043704,PhysRevLett.103.077207,PhysRevB.83.180407,PhysRevLett.114.227202,1602.04028} The small energy scale for the magnetic order and unusual response to an external magnetic field make it difficult to identify the magnetic order below $1 \, \mathrm{K}$ and may also suggest that a number of exotic quantum ground states may arise in this system. On the other hand, a recent thermal conductivity measurement finds signatures of entropy-carrying charge-neutral excitations above $1 \, \mathrm{K}$, which suggests that the phase above $1 \, \mathrm{K}$ may be connected to a putative quantum spin liquid with spinon quasiparticles.\cite{1608.00444} Over the years, Volborthite has also inspired several theoretical studies.\cite{PhysRevB.76.064430,PhysRevB.76.094421,PhysRevB.78.174420,JPSJ.79.073708,1610.03135}

In this work, we investigate possible quantum spin liquid and magnetically ordered phases on the distorted Kagome lattice, in view of the spatially anisotropic spin exchange model proposed earlier by the density functional theory (DFT) computations.\cite{PhysRevLett.117.037206} The distorted Kagome lattice corresponding to the so-called $J-J'-J_1-J_2$ model represents the non-symmorphic layer space group, $p2gg$, that possesses glide symmetry, the combination of reflection and a fractional translation. Motivated by the experiments and the DFT result, we first classify possible spin liquid phases with bosonic spinons for the $p2gg$ space group via the projective symmetry group (PSG) analysis \cite{PhysRevB.65.165113} of the Schwinger boson \cite{PhysRevB.45.12377,PhysRevB.74.174423,PhysRevB.87.125127} mean field states. It is shown that only the translationally invariant mean field states are possible in two-dimensional non-symmorphic lattices. As a result, there are only eight possible spin liquid phases characterized by three $\mathbb{Z}_2$ variables when both of the spatial and time reversal symmetries are taken into account. Using these results, we investigate the mean field ground states of the $J-J'-J_1-J_2$ model. We identify the stable quantum spin liquid states and obtain magnetically ordered phases that arise from these spin liquid states by condensing the bosonic spinons.

It is found that the spin liquid ground state of this model is connected to a coplanar incommensurate $(q,0)$ spiral order in the semiclassical limit while a different, but highly competing, spin liquid state is related to a coplanar commensurate $(\pi,\pi)$ spin density wave phase, where the amplitude of the spin density varies from site to site. The corresponding classical model is also studied by simulated annealing and we find that the classical magnetic order is indeed consistent with the $(q,0)$ spiral order for the model parameters determined by the DFT computations. This correspondence is natural as the length of classical spin is fixed and hence the competing spin density wave order is simply not possible in the classical model. Interestingly, the spin liquid state related to the $(\pi,\pi)$ spin density wave order exhibits periodicity enhancement, namely an extra periodicity beyond the normal periodicity given by the lattice structure, in the two-spinon continuum. 
Such periodicity enhancement is a signature of the so-called symmetry fractionalization,\cite{PhysRevB.87.104406,PhysRevB.90.121102,1403.0575} which originates from the existence of spatial inversion and time reversal symmetry. 

We argue that it may be necessary to consider both spin liquid states described above on equal footing when we apply these results to experiments. This is because the mean field energetics may not be accurate enough to determine the true ground state and the model parameters determined by the DFT computations may also allow some variations. In fact, both of the incommensurate $(q.0)$ spiral order and the commensurate $(\pi,\pi)$ spin density wave order may be compatible with the existing nuclear magnetic resonance (NMR) experimental data on Volborthite below 1K. An important question is whether the paramagnetic state above 1K can be regarded as a continuation of certain quantum spin liquid state, as suggested by the recent thermal conductivity measurement. In view of this possibility, it will be interesting to perform a neutron scattering experiment on Volborthite above 1K and look for signatures of the two-spinon continuum and especially periodicity enhancement of such continuum of excitations.

The rest of this paper is organized as follows. In Section \ref{latticehamiltonian}, we introduce the lattice structure that is compatible with the spin model derived from DFT computations. We explain the non-symmorphic nature of this lattice and some details of the DFT-derived spin model. In Section \ref{schwingerbosonmeanfieldtheory}, we outline the Schwinger boson mean field theory. In Section \ref{bosonicpsganalysis}, we perform the PSG analysis of quantum spin liquid phases with bosonic spinons and construct relevant mean field ansatz for the spin liquid states. In Section \ref{resultdiscuss}, we analyze the energetics of the mean field spin liquid states via Schwinger boson mean field theory and the two-spinon continuum in each spin liquid state. We investigate the magnetically ordered states obtained by condensing bosonic spinons in the spin liquid phases. We then compare these results with the simulated annealing study of the corresponding classical model. 
In Section \ref{summary}, we discuss relevance of our results to the existing and future experiments on Volborthite.

\section{\label{latticehamiltonian} Lattice Structure and Spin Model}
In Volborthite $\mathrm{Cu_3 V_2 O_7 (OH)_2 \cdot 2 H_2 O}$, two layers of distorted Kagome lattice, each consisting of edge-sharing $\mathrm{CuO_6}$ octahedra, are separated by non-magnetic $\mathrm{V_2 O_7}$ pillars and $\mathrm{H_2O}$ molecules.\cite{ncomms1875} It is reasonable to assume that the interaction between different Kagome layers is negligible and we will therefore focus on just a single Kagome layer. The localized $S=1/2$ moment at each site on the Kagome lattice is carried by $\mathrm{Cu^{2+}}$ ions.\cite{ncomms1875} Moreover, there are two crystallographically distinct $\mathrm{Cu^{2+}}$ sites, which suggests two different magnetically active orbitals.\cite{PhysRevLett.117.037206} As shown below, the structure of the distorted Kagome layer is described by the plane crystallographic group $p2gg$, whose space group is discussed in Section \ref{unitcellspacegroup}. This non-symmorphic version of the Kagome lattice (non-symmorphic Kagome lattice hereafter) possesses glide symmetry. We introduce the microscopic spin model of Volborthite, derived from a recent density functional theory (DFT) calculation \cite{PhysRevLett.117.037206} in Section \ref{microscopic}.

\subsection{\label{unitcellspacegroup} Unit Cell and Space Group}
\begin{figure}
\includegraphics[scale=0.3]{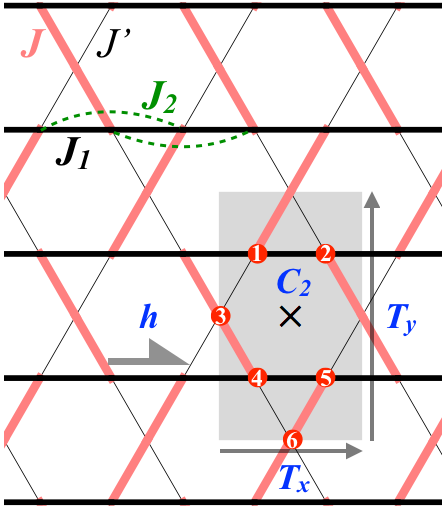}
\caption{\label{unitcellspacegroupfigure} 
The lattice structure of the non-symmorphic Kagome lattice, and the spin model for Volborthite obtained from DFT calculation. There are six sites (numbered circles) per unit cell (shaded region).
The space group elements we consider are lattice translations $T_x$ and $T_y$, $\pi$-rotation $C_2$, and glide $h$.
The four leading exchange interactions in Volborthite are given by $J:J':J_1:J_2=1:-0.2:-0.5:0.2$. $J_1$ describes the first nearest neighbour coupling in the chain direction, for example between $(x,y,1)$ and $(x,y,2)$. $J_2$ describes the second nearest neighbour coupling in the chain direction, for example between $(x,y,1)$ and $\left(x+1,y,1\right)$. $J$ and $J'$ describe two different couplings between a chain site and an interstitial site, for example $(x,y,3)$ with $(x,y,4)$ and $(x,y,1)$ with $(x,y,3)$ respectively.
}
\end{figure}

The non-symmorphic Kagome lattice has six sites (or sublattices) per unit cell (FIG. \ref{unitcellspacegroupfigure}). Denote the lattice constant along $x$- and $y$-direction by $b$ and $a$ respectively. Then, the coordinate of a generic site has the form $(xb,ya,s)$, which we simply write as $(x,y,s)$, where $x,y \in \mathbb{Z}$ specify the unit cell to which the site belongs, and $s=1,\ldots,6$ indexes the sublattice. The space group of non-symmorphic Kagome lattice is generated by $\pi$-rotation $C_2$ and glide $h$, which consists of reflection and half lattice translation. In general, a non-symmorphic operation combines a point group operation (e.g. rotation and reflection) with a fractional lattice translation, which cannot be rewritten in terms of point group operations and full lattice translations by switching to another coordinate system.\cite{nphys2600} The non-symmorphic symmetry has important implication on the translational invariance of mean field ansatzes, which is discussed in Section \ref{meanfieldansatzboson}.

For convenience of subsequent analysis, we also consider the lattice translations $T_x$ and $T_y$ along two independent directions $\hat{\mathbf{x}}$ and $\hat{\mathbf{y}}$. We fix the center of rotation at the center of hexagon in the $(0,0)$ unit cell, and the glide axis to the horizontal line passing through $(0,0,4)$ and $(0,0,5)$. $h$ is therefore the reflection about the glide axis followed by translation by $b\hat{\mathbf{x}}/2$. In Appendix \ref{spacegroupalgebra} we show explicitly how a site with coordinates $(x,y,s)$ transforms under the space group operations.
We refer to the direction along sublattices $\ldots -1-2-1-2- \ldots$ (equivalently $\ldots -4-5-4-5- \ldots$ as they are related by symmetry) as the chain direction. The sites between adjacent chains, which have either sublattice index $3$ or $6$, are known as interstitial sites.

\subsection{\label{microscopic} Microscopic Spin Model}
According to the DFT calculation by Janson \textit{et al},\cite{PhysRevLett.117.037206} the interaction between $S=1/2$ local moments on the Volborthite Kagome layer is described by the following Heisenberg model
\begin{equation} \label{heisenberggeneral}
H = \sum_{ij} J_{ij} \mathbf{S}_i \cdot \mathbf{S}_j
\end{equation}
with four leading exchange interactions. They are the first and second nearest neighbour couplings along the chain direction, $J_1$ and $J_2$, and two different couplings between the chain and interstitial spins, $J$ and $J'$, as shown in FIG. \ref{unitcellspacegroupfigure}. The ratio between these coupling constants is given by $J:J':J_1:J_2=1:-0.2:-0.5:0.2$, with negative (positive) sign indicating ferromagnetic (antiferromagnetic) interaction.

\section{\label{schwingerbosonmeanfieldtheory} Schwinger Boson Mean Field Theory}
In this section, we outline the Schwinger boson mean field approach \cite{PhysRevB.45.12377,PhysRevB.74.174423,PhysRevB.87.125127} for the generic Heisenberg Hamiltonian \eqref{heisenberggeneral}. We start with Schwinger boson representation of the spin operator,
\begin{equation} \label{spinoperatorboson}
\mathbf{S}_i = \frac{1}{2} \sum_{\alpha \beta} b_{i \alpha}^\dagger \bm{\sigma}_{\alpha \beta} b_{i \beta} ,
\end{equation}
where $b_{i\alpha}^{\dagger}$ ($b_{i \alpha}$) creates (annihilates) a bosonic spinon with spin $\alpha \in \lbrace \uparrow , \downarrow \rbrace$ at site $i$. Here $\bm{\sigma}=(\sigma^x,\sigma^y,\sigma^z)$ is the vector of Pauli matrices. The bosonic operators obey commutation relation,
\begin{equation} \label{commutation}
\begin{aligned}
\left[b_{i \alpha},b_{j \beta}^\dagger\right] &= \delta_{ij} \delta_{\alpha \beta} , \\
\left[b_{i \alpha},b_{j \beta}\right] &= 0 = \left[b_{i \alpha}^\dagger,b_{j \beta}^\dagger\right] .
\end{aligned}
\end{equation}
The total number of Schwinger bosons at site $i$ is represented by the number operator
\begin{equation} \label{numberoperatorboson}
\hat{n}_i = \sum_\alpha b_{i \alpha}^\dagger b_{i \alpha}.
\end{equation}
On the other hand, for localized moments of spin $S$, the total spin operator reads \cite{PhysRevB.87.125127}
\begin{equation} \label{totalspin}
\begin{aligned}[b]
\mathbf{S}_i^2 &= S\left(S+1\right) \\
&= \frac{\hat{n}_i}{2} \left(\frac{\hat{n}_i}{2} + 1 \right) ,
\end{aligned}
\end{equation}
where the second equality follows from \eqref{spinoperatorboson} and \eqref{numberoperatorboson}. This imposes a constraint on the number operator,
\begin{equation} \label{numberconstraintboson}
\hat{n}_i = 2S \equiv \kappa .
\end{equation}
Following Refs.~\onlinecite{PhysRevB.74.174423,PhysRevB.45.12377}, we allow $\kappa$ to be any positive real number, i.e. $S$ is not restricted to $1/2$. At the mean field level, the constraint is replaced by its ground state expectation value,\cite{PhysRevB.74.174423}
\begin{equation} \label{numberconstraintbosonaverage}
\langle \hat{n}_i \rangle = \kappa .
\end{equation}

Next, we define the bond operators \cite{PhysRevB.74.174423}
\begin{subequations}
\begin{align}
\hat{A}_{ij} &= \frac{1}{2} \sum_{\alpha \beta} b_{i \alpha} \epsilon_{\alpha \beta} b_{j \beta}, \label{pairingoperatorboson} \\
\hat{B}_{ij} &= \frac{1}{2} \sum_{\alpha} b_{i \alpha}^\dagger b_{j \alpha} . \label{hoppingoperatorboson}
\end{align}
\end{subequations}
with $\epsilon_{\alpha \beta}$ being the antisymmetric tensor. $\hat{A}_{ij}$ ($\hat{B}_{ij}$) is known as singlet pairing (hopping) channel. It can be checked explicitly that both $\hat{A}_{ij}$ and $\hat{B}_{ij}$ are invariant under global $SU(2)$ spin rotation. In terms of bond operators, the dot product of spin operators can be written as
\begin{equation} \label{spindotproductboson}
\mathbf{S}_i \cdot \mathbf{S}_j = :\hat{B}_{ij}^\dagger \hat{B}_{ij}: - \hat{A}_{ij}^\dagger \hat{A}_{ij} .
\end{equation}
Heisenberg Hamiltonian \eqref{heisenberggeneral} then becomes
\begin{equation} \label{heisenbergpairinghopping}
H = \sum_{ij} J_{ij} \left( :\hat{B}_{ij}^\dagger \hat{B}_{ij}: - \hat{A}_{ij}^\dagger \hat{A}_{ij} \right) - \sum_i \mu_i \left( \hat{n}_i - \kappa \right) .
\end{equation}
We have imposed the contraint \eqref{numberconstraintboson} on all sites by introducing Lagrange multipliers (chemical potentials) $\mu_i$ in the Hamiltonian. Here $: \; :$ denotes normal ordering.

The Hamiltonian \eqref{heisenbergpairinghopping} is quartic in bosonic operators $b$. We perform mean field decoupling on $\hat{B}_{ij}^\dagger \hat{B}_{ij}$ and $\hat{A}_{ij}^\dagger \hat{A}_{ij}$ to obtain a Hamiltonian quadratic in $b$.\cite{PhysRevB.74.174423} Such a decoupling preserves the global $SU(2)$ spin rotation symmetry, which is present in the original Heisenberg Hamiltonian \eqref{heisenberggeneral}. Then, we replace $\langle \hat{A}_{ij} \rangle$ and $\langle \hat{B}_{ij} \rangle$ by complex-valued variational parameters $A_{ij}$ and $B_{ij}$ respectively. The mean field Hamiltonian reads
\begin{equation} \label{heisenbergmeanfieldpairinghoppingboson}
\begin{aligned}[b]
H_{\mathrm{MF}} =& \sum_{ij} J_{ij} \left[ \left( {B}_{ij}^* \hat{B}_{ij} + \mathrm{h.c.} - \left\lvert B_{ij} \right\rvert^2 \right) \right. \\
& \left. \qquad \quad \; - \left( {A}_{ij}^* \hat{A}_{ij} + \mathrm{h.c.} - \left\lvert A_{ij} \right\rvert^2 \right) \right] \\
& - \sum_i \mu_i \left( \hat{n}_i - \kappa \right) .
\end{aligned}
\end{equation}
Extremizing the expectation value of mean field Hamiltonian with respect to the variational parameters yields the following self-consistent equations,\cite{PhysRevB.74.174423}
\begin{equation}
\frac{\partial \langle H_\mathrm{MF} \rangle}{\partial \mathcal{O}_{ij}} = 0 \iff \mathcal{O}_{ij} = \langle \hat{\mathcal{O}}_{ij} \rangle, \, \mathcal{O}_{ij}=A_{ij}, B_{ij} , \label{selfconsistentequationboson}
\end{equation}
and the chemical potential yields the constraint \eqref{numberconstraintbosonaverage},
\begin{equation}
\frac{\partial \langle H_\mathrm{MF} \rangle}{\partial \mu_i} = 0 \iff \kappa = \langle \hat{n}_i \rangle . \label{selfconsistentequationchemical}
\end{equation}

In the mean field decoupling scheme above, however, $A_{ij}$ ($B_{ij}$) channel would present an unbounded free energy for $J_{ij} < 0$ ($J_{ij} > 0$). In order to obtain a controlled mean field solution, we use the following identity \cite{PhysRevB.87.125127}
\begin{equation} \label{pairinghoppingidentity}
:\hat{B}_{ij}^\dagger \hat{B}_{ij}: + \hat{A}_{ij}^\dagger \hat{A}_{ij} = \frac{1}{4} \hat{n}_i \left( \hat{n}_j - \delta_{ij} \right) ,
\end{equation}
which leads to two variants of \eqref{spindotproductboson},
\begin{subequations} \label{spindotproductbosonvariant}
\begin{align}
\mathbf{S}_i \cdot \mathbf{S}_j &= 2 :\hat{B}_{ij}^\dagger \hat{B}_{ij}: - S^2 \label{spindotproductbosonhopping} \\
&= S^2 - 2 \hat{A}_{ij}^\dagger \hat{A}_{ij} , 
\label{spindotproductbosonpairing}
\end{align}
\end{subequations}
where we have assumed the constraint \eqref{numberconstraintboson} and $i \neq j$. Therefore, for $J_{ij}>0$ ($J_{ij}<0$), we write the spin scalar product as \eqref{spindotproductbosonpairing} (\eqref{spindotproductbosonhopping}), which contains only $A_{ij}$ ($B_{ij}$). The multiplicative constant $2$ and the additive constant $\pm S^2$ can be further dropped without qualitatively changing the theory. Mean field decoupling then leads to the following Hamiltonian
\begin{equation} \label{heisenbergmeanfieldboson}
\begin{aligned}[b]
H_{\mathrm{MF}} = & - \sum_{J_{ij}>0} J_{ij} \left( {A}_{ij}^* \hat{A}_{ij} + \mathrm{h.c.} - \left\lvert A_{ij} \right\rvert^2 \right) \\
& + \sum_{J_{ij}<0} J_{ij} \left( {B}_{ij}^* \hat{B}_{ij} + \mathrm{h.c.} - \left\lvert B_{ij} \right\rvert^2 \right) \\
& - \sum_i \mu_i \left( \hat{n}_i - \kappa \right)
\end{aligned}
\end{equation}
while the self-consistent equations are given by the same expression \eqref{selfconsistentequationboson} as before. In addition, we assume that symmetry-related sites have the same chemical potential. Since there exist  two chemically inequivalent sites, we denote the chemical potential at the chain sites ($s=1,2,4,5$) by $\mu_1$, and the interstitial sites ($s=3,6$) by $\mu_2$. The constraint \eqref{numberconstraintbosonaverage} is then split into two,
\begin{subequations}
\begin{align}
\frac{1}{4} \sum_{s=1,2,4,5} \langle \hat{n}_{i,s} \rangle &= \kappa , \label{numberconstraintboson1245} \\
\frac{1}{2} \sum_{s=3,6} \langle \hat{n}_{i,s} \rangle &= \kappa , \label{numberconstraintboson36}
\end{align}
\end{subequations}
where $i$ now labels the unit cells rather than the individual sites. Notice that $\kappa$ is still the same at both sites since all spins have the same amplitude.

In practice, the self-consistent equations \eqref{selfconsistentequationboson} are solved iteratively in momentum space. The procedure and techniques of mean field calculation are discussed in Appendix \ref{meanfieldcalculation}.

\section{\label{bosonicpsganalysis} Projective Symmetry Group Analysis}
We first classify possible $\mathbb{Z}_2$ spin liquid phases with bosonic spinons for the two-dimensional non-symmorphic lattice described by $p2gg$ space group. Notice that this classification is independent of the spin Hamiltonian. If a specific spin model is chosen, one can investigate the spin liquid ground state that minimizes the mean field energy. We use the projective symmetry group (PSG) analysis,\cite{PhysRevB.65.165113,PhysRevB.74.174423,PhysRevB.87.125127} namely we classify possible spin liquid mean field ansatzes that are invariant under the lattice symmetry transformations followed by a gauge transformation. Recall that the global spin rotation symmetry is automatically enforced by the choice of mean field parameters $A_{ij}$ and $B_{ij}$. Hence, in the following, we consider the space group and time reversal symmetry for the PSG analysis.

First, we observe that the mean field Hamiltonian \eqref{heisenbergmeanfieldboson} has $U(1)$ gauge redundancy. It is invariant under the following local $U(1)$ transformation \cite{PhysRevB.74.174423,PhysRevB.87.125127}
\begin{subequations}
\begin{align}
G: \: & b_{i \alpha} \longrightarrow e^{i\phi(i)} b_{i \alpha} , \label{gaugetransformationboson} \\
& A_{ij} \longrightarrow e^{i\left(\phi(i)+\phi(j)\right)} A_{ij} , \label{gaugetransformationpairingansatz} \\
& B_{ij} \longrightarrow e^{i\left(-\phi(i)+\phi(j)\right)}B_{ij} . \label{gaugetransformationhoppingansatz}
\end{align}
\end{subequations}
Notice that the gauge transformed bosonic operators $\tilde{b}_{i \alpha} \equiv e^{i\phi(i)} b_{i \alpha}$ leave the representation \eqref{spinoperatorboson} of spin invariant, satisfy the same commutation relation \eqref{commutation} and constraint \eqref{numberconstraintboson}, so that it describes the same physical spin. This implies that mean field ansatzes differing by a gauge transformation would correspond to the same physical state.

Now let us consider a space group element $X$ acting on the bosonic operator via
\begin{equation} \label{symmetrytransformationboson}
X: b_{i \alpha} \longrightarrow b_{X(i) \alpha} .
\end{equation}
Instead of considering the action on the bosonic operator, we can equivalently view $X$ as acting on the ansatz. We want $H_{\mathrm{MF}}$ to respect every symmetry operation $X$, such that the physics it describes is invariant under the transformation \eqref{symmetrytransformationboson}. Therefore, under the action of $X$, the modified ansatz is equivalent to the original one up to a gauge transformation, because they result in the same physical state. The set of all compound operations consisting of symmetry and gauge transformations that leave the ansatz invariant is defined as the \textit{projective symmetry group} (PSG).\cite{PhysRevB.65.165113} To illustrate this, consider for example the term $A_{ij} \hat{A}_{ij}^\dagger \sim A_{ij} b_{i \alpha}^\dagger b_{j \beta}^\dagger$ in the mean field Hamiltonian \eqref{heisenbergmeanfieldboson}. Applying the symmetry transformation $X$ followed by the associated gauge transformation $G_X(\mathbf{r}) = \exp\left[i\phi_X(\mathbf{r})\right]$ on such a term, we get \cite{PhysRevB.87.235103}
\begin{equation*}
\begin{aligned}
A_{ij} b_{i \alpha}^\dagger b_{j \beta}^\dagger & \overset{X}{\longrightarrow} A_{ij} b_{X(i) \alpha}^\dagger b_{X(j) \beta}^\dagger \\
& \overset{G_X}{\longrightarrow} A_{ij} e^{-i\left(\phi_X(X(i))+\phi_X(X(j))\right)} b_{X(i) \alpha}^\dagger b_{X(j) \beta}^\dagger .
\end{aligned}
\end{equation*}
With the sites $i$ and $j$ being summed over, the invariance of the mean field ansatz and thus $H_\mathrm{MF}$ under $G_X X$ requires that
\begin{equation} \label{pairingbosonpsgrelation}
A_{ij} e^{-i\left(\phi_X(X(i))+\phi_X(X(j))\right)} = A_{X(i)X(j)} ,
\end{equation}
or $A_{ij}=\exp \left[i\phi_X(X(i))+i\phi_X(X(j))\right] A_{X(i)X(j)}$ as \eqref{gaugetransformationpairingansatz} and \eqref{symmetrytransformationboson} might intuitively suggest. It can be similarly shown that
\begin{equation} \label{hoppingbosonpsgrelation}
B_{ij} e^{-i\left(-\phi_X(X(i))+\phi_X(X(j))\right)} = B_{X(i)X(j)} .
\end{equation}
Then, the collection of all $G_XX$ that leave the ansatz invariant is called the PSG. This definition also includes $G_\mathcal{T} \mathcal{T}$ for the antiunitary time reversal $\mathcal{T}$. We discuss how to treat time reversal symmetry explicitly in bosonic PSG in Appendix \ref{timereversalpsgboson}.

Suppose that $G_X X \in \mathrm{PSG}$ for a space group element $X$. If we apply a gauge transformation on the mean field ansatz, say $A_{ij} \longrightarrow G A_{ij}$, then $G_X \longrightarrow G G_X X G^{-1} X^{-1}$ such that $G_X X$ is still an element of the PSG.\cite{PhysRevB.74.174423} The corresponding change in the phase of the bosonic operator is given by
\begin{equation} \label{gaugetransformpsgspaceboson}
\phi_X (\mathbf{r}) \longrightarrow \phi (\mathbf{r}) + \phi_X (\mathbf{r}) - \phi (X^{-1}(\mathbf{r})) .
\end{equation}
Due to the antiunitarity of $\mathcal{T}$, $G_\mathcal{T}$ transforms in a different manner \eqref{gaugetransformpsgtimeboson}, which is explained in Appendix \ref{timereversalpsgboson}.

The PSG elements of the form $G_II$ with $I$ being the identity element of the space group form a subgroup of PSG called the invariant gauge group (IGG). \cite{PhysRevB.87.125127} Alternatively, we can view IGG as the set of all pure gauge transformations that leave the ansatz invariant. For the mean field Hamiltonian \eqref{heisenbergmeanfieldboson} in which both $A_{ij}$ and $B_{ij}$ are present, the IGG is just $\mathbb{Z}_2=\lbrace -1, 1 \rbrace$. \cite{PhysRevB.74.174423}

\subsection{\label{algebraicpsgboson} Algebraic PSG}
The algebraic relations among space group elements constrain the possible forms of gauge transformations $G_X$. \cite{PhysRevB.74.174423} For instance, consider the string of translation operators that equals to identity,
\begin{equation} \label{algebraicidentityexample}
T_x^{-1} T_y^{-1} T_x T_y = I .
\end{equation}
Suppose $G_{T_x}T_x,G_{T_y}T_y \in \mathrm{PSG}$, then we must have
\begin{equation} \label{algebraicconditionexample}
\left(G_{T_x}T_x\right)^{-1} \left(G_{T_y}T_y\right)^{-1} \left(G_{T_x}T_x\right) \left(G_{T_y}T_y\right) \in \mathrm{IGG} ,
\end{equation}
or, in terms of phases,
\begin{equation*}
\begin{aligned}
& - \phi_{T_x}(x+1,y,s) - \phi_{T_y}(x+1,y+1,s) \\
& \qquad + \phi_{T_x}(x+1,y+1,s) + \phi_{T_y}(x,y+1,s) = n \pi, \, n=0,1 .
\end{aligned}
\end{equation*}
The PSG in which the gauge transformations $G_X$ satisfy algebraic constraints such as \eqref{algebraicconditionexample} is called algebraic PSG. For a given lattice, there is only a finite number of independent algebraic identities such as \eqref{algebraicidentityexample}. These identities can be found by inspecting how two distinct space group elements commute, which are listed in Appendix \ref{spacegroupalgebra} in the case of the non-symmorphic Kagome lattice. Representing the gauge transformations by their phases, the final solution of algebraic PSG is given by 
\begin{subequations}
\begin{align}
\phi_{T_x}\left(x,y,s\right) &= 0 , \label{translationxpsgboson} \\
\phi_{T_y}\left(x,y,s\right) &= 0 , \label{translationypsgboson} \\
\phi_{C_2}\left(x,y,s\right) &= \phi_{C_2}\left(0,0,s\right) + p_3 \pi \left( x + y \right) , \\
\phi_h\left(x,y,s\right) &= \phi_h\left(0,0,s\right) + p_3 \pi y , \\
\phi_\mathcal{T}\left(x,y,s\right) &= \phi_\mathcal{T}\left(0,0,s\right) ,
\end{align}
\end{subequations}
with
\begingroup
\allowdisplaybreaks
\begin{align*}
\phi_{C_2} (0,0,s=1,2) & = 0 , \\
\phi_{C_2} (0,0,s=4,5) & = p_2 \pi , \\
\phi_{C_2} (0,0,s=3) & = \frac{p_2+p_3}{2} \pi , \\
\phi_{C_2} (0,0,s=6) & = \frac{3\left(p_2+p_3\right)}{2} \pi , \\
\phi_h (0,0,s=1) & = p_2 \pi , \\
\phi_h (0,0,s=2) & = \left(p_2+p_3\right) \pi , \\
\phi_h (0,0,s=3,4,5,6) & = 0 , \\
\phi_\mathcal{T}\left(0,0,s=1,3\right) &= 0 , \\
\phi_\mathcal{T}\left(0,0,s=2,6\right) &= p_{13}\pi , \\
\phi_\mathcal{T}\left(0,0,s=4\right) &= \left(p_2 + p_3 + p_{13}\right)\pi , \\
\phi_\mathcal{T}\left(0,0,s=5\right) &= \left(p_2 + p_3\right)\pi .
\end{align*}
\endgroup
The three independent $\mathbb{Z}_2$ variables $p_2, p_3, p_{13} \in \lbrace 0, 1 \rbrace$ lead to $2^3=8$ distinct bosonic spin liquid states. Detailed derivation of the algebraic PSG can be found in Appendix \ref{bosonicpsgderive}. We remark that, while $p_2$ and $p_3$ arise entirely from the space group considerations, $p_{13}$ is introduced only when time reversal symmetry/invariance is explicitly enforced. If only the spatial symmetry is enforced, the spin liquid phases are simply classified by $(p_2,p_3)$, which would in principle allow both time reversal invariant as well as time reversal breaking spin liquid states. In the mean field analysis of spin liquid phases in Section \ref{resultdiscuss}, we will study possible spin liquid states by requiring only the spatial symmetry. However, we will find that all of the stable mean field solutions, which belong to the $(p_2,p_3)$ classification, satisfy time reversal symmetry and they are related to the time reversal invariant $(p_1,p_2,p_{13}=0)$ states. 

\subsection{\label{meanfieldansatzboson} Mean Field Ansatz}
Considering the spin model described in Section \ref{microscopic} and the generic mean field Hamiltonian \eqref{heisenbergmeanfieldboson}, we can see that there are altogether four independent pairing and hopping amplitudes per unit cell, which we denote by $A$, $B'$, $B_1$ and $A_2$, depending on which exchange coupling they are associated with. All other amplitudes can be generated from these by lattice symmetry transformations. For example, let us fix $A=A_{(0,0,3) \longrightarrow (0,0,4)}$. The amplitude $A_{\left(1,0,3\right) \longrightarrow (0,0,2)}$ is related to $A_{(0,0,3) \longrightarrow (0,0,4)}$ by $C_2$ and the PSG ensures that the mean field ansatz is invariant under $G_{C_2}C_2$. Using \eqref{pairingbosonpsgrelation},
\begin{equation} \label{pairingrelationbosonexample}
\begin{aligned}[b]
A_{\left(1,0,3\right) \longrightarrow (0,0,2)}
&= A_{C_2(0,0,3) \longrightarrow C_2(0,0,4)} \\
&= e^{-i\left(\phi_{C_2}(1,0,3)+\phi_{C_2}(0,0,2)\right)} A_{(0,0,3) \longrightarrow (0,0,4)} \\
&= e^{-i \left(p_2+3p_3\right) \pi/2} A .
\end{aligned}
\end{equation}
If time reversal symmetry is considered, the corresponding phase $\phi_\mathcal{T}$ further constrains the complex phase of $A$, $B'$, $B_1$ and $A_2$ (see Appendix \ref{timereversalpsgboson}).

Notice that the phase variables related to the lattice translations, as shown in \eqref{translationxpsgboson} and \eqref{translationypsgboson}, are trivial. To construct $H_\mathrm{MF}$, it is therefore sufficient to determine various relations between mean field amplitudes such as \eqref{pairingrelationbosonexample} in the $\left(0,0\right)$ unit cell, as all other unit cells have the same relations via \eqref{pairingbosonpsgrelation} and \eqref{hoppingbosonpsgrelation} with $X=T_x,T_y$. In other words, the mean field ansatz does not go beyond the physical unit cell. This is a consequence of the non-symmorphic symmetry of the lattice, which we explain as follows. The algebraic identities \eqref{algebraicrelation5} and \eqref{algebraicrelation7} impose the constraints ${\tilde{T}_x}^{-1} \tilde{h}^2 = \pm 1 \equiv \eta_{h}$ and $\tilde{h}^{-1} \tilde{T}_y \tilde{h} \tilde{T}_y= \pm 1 \equiv \eta_{h T_y}$ on PSG, where we have used the abbreviation $\tilde{X} = G_X X$. Then,
\begin{equation} \label{translationfractionalization}
\begin{aligned} [b]
{\tilde{T}_x}^{-1} {\tilde{T}_y}^{-1} \tilde{T}_x \tilde{T}_y &= (\eta_{h} \tilde{h}^2)^{-1} {\tilde{T}_y}^{-1} \eta_{h} \tilde{h}^2 \tilde{T}_y \\
&= \tilde{h}^{-1} \left(\tilde{h}^{-1} {\tilde{T}_y}^{-1} \tilde{h} {\tilde{T}_y}^{-1}\right) \tilde{T}_y \tilde{h} \tilde{T}_y \\
&= \eta_{hT_y} \eta_{hT_y} \\
&= +1 .
\end{aligned}
\end{equation}
\eqref{translationfractionalization} dictates that the mean field ansatz constructed from PSG (see Section \ref{meanfieldansatzboson}) can never enlarge the physical unit cell, because it would require ${\tilde{T}_x}^{-1} {\tilde{T}_y}^{-1} \tilde{T}_x \tilde{T}_y=-1$ if the contrary were true. This argument also holds for three other non-symmorphic plane crystallographic groups $pg$, $p2mg$ and $p4gm$. Such feature is characteristically different from the isotropic Kagome lattice, where, for example, the glide symmetry is absent. The algebraic PSG of the isotropic Kagome lattice allows certain spin liquid states such as the $\pi$-flux state (in the terminology of Ref.~\onlinecite{PhysRevB.74.174423}) that is given by a mean field ansatz that enlarges the physical unit cell.

\section{\label{resultdiscuss} Emergent Quantum Phases and Their Physical Properties}

As mentioned earlier, we consider the spin liquid phases labeled by $(p_2,p_3)$, which are constrained from the spatial symmetries.
We determine the ground state of the spin model derived in the DFT computation, for different values of $2S = \kappa$. In Schwinger boson mean field theory, the condensation of spinons at certain wavevector arises as the number of bosons per site, $\kappa$, increases and then exceeds a critical value $\kappa_c$.\cite{PhysRevB.74.174423} This corresponds to the transition from a given spin liquid state to a magnetically ordered state. Notice that increasing the ``spin" magnitude $S = \kappa /2$ reduces quantum fluctuations. The magnetically ordered phases that arise right after the transition can be obtained by analyzing the critical eigenmode near the transition. We identify such magnetically ordered phases obtained from different spin liquid phases. We also compare these results to the simulated annealing result of the classical model, which is equivalent to $\kappa \longrightarrow \infty$ limit in the Schwinger boson formulation.

In order to better characterize the spin liquid states, we compute the dispersion of the lower boundary of two-spinon continuum.
As shown below, the spin liquid phases labeled by $p_3=1$ exhibit periodicity enhancement, namely the two-spinon continuum are invariant under translation by $\mathbf{k}=(\pm \pi,\pm \pi)$ in momentum space, which leads to spectral doubling in the Brillouin zone. We make the connection between this phenomenon and the notion of symmetry fractionalization \cite{PhysRevB.87.104406,PhysRevB.90.121102} introduced in previous works.

\subsection{\label{bosonicspinliquid} Spin Liquid States}

Consider the spin liquid states labeled by $(p_2,p_3)$, where $p_2,p_3 \in \lbrace 0,1 \rbrace$. The critical $\kappa=\kappa_c$, where the spinons condense and the minimum of spinon dispersion $\omega_\mathrm{min}$ touches zero energy, is shown in TABLE \ref{densityandstate}. When $\kappa > \kappa_c$, a magnetic order arises. The value of $\kappa_c$ is obtained by computing $\omega_{\rm min}$ at various values of $\kappa$ and then making a linear extrapolation, as shown in FIG. \ref{linearapproximationfigure} in Appendix \ref{linearapproximationscheme}. The values of $\kappa_c$ fall between 0.3 and 0.4.

We compare the total energy $E=\langle H_\mathrm{MF} \rangle$ of four spin liquid states at several values of $\kappa<\kappa_c$, as shown in TABLE \ref{energycomparisontable} in Appendix \ref{energycomparison}. At $\kappa$ far below $\kappa_c$, the spin liquid states with the same $p_2$ have exactly the same energy, which happens because each of the hopping amplitudes $B_1$ and $B'$ vanishes while each of the pairing amplitudes $A_2$ and $A$ has the same magnitude. As $\kappa$ approaches $\kappa_c$, the situation changes and all of these amplitudes become finite, which lifts the degeneracy. We find that $(p_2,p_3)=(1,0)$ is the most energetically favorable state among four spin liquid phases. Nevertheless, it should be noticed that the energy of $(1,1)$ spin liquid state is particularly close to $(1,0)$. We will pay particular attention to $(1,0)$ and $(1,1)$ spin liquid states as they are closely competing phases.

\begin{table}
\caption{\label{densityandstate} Critical bosonic density $\kappa_c$ of different spin liquid states $\left(p_2,p_3\right)$.}
\begin{tabular}{>{\centering\arraybackslash} m{2 cm}|>{\centering\arraybackslash} m{2 cm}}
\hline \hline
$\left(p_2,p_3\right)$ & $\kappa_\mathrm{c}$ \\ \hline
$\left(0,0\right)$ & $0.327$ \\
$\left(1,0\right)$ & $0.351$ \\
$\left(0,1\right)$ & $0.362$ \\
$\left(1,1\right)$ & $0.368$ \\
\hline \hline
\end{tabular}
\end{table}

\begin{figure*}
\begin{tabular}{>{\centering\arraybackslash} m{6.8 cm} >{\centering\arraybackslash} m{0.25 cm} >{\centering\arraybackslash} m{6.8 cm}}
\includegraphics[scale=0.32]{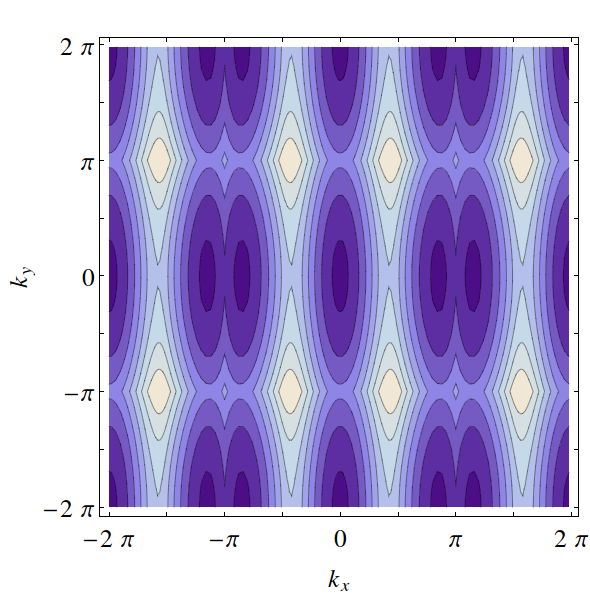} & & \includegraphics[scale=0.32]{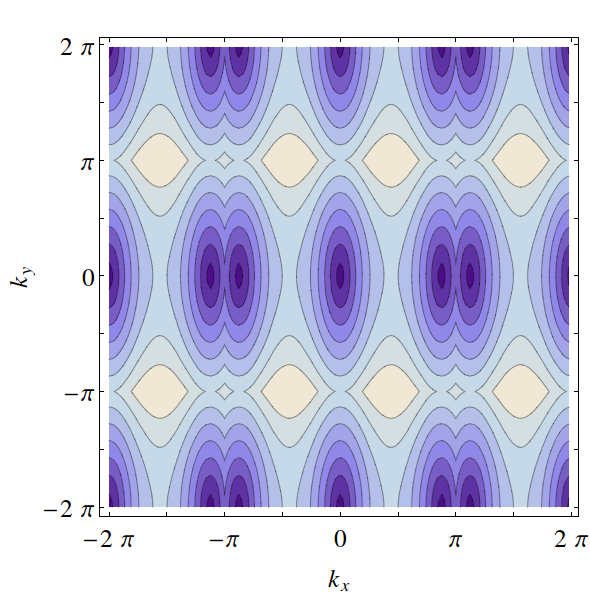} \\
\qquad \quad \ (a) \ \ $(p_2,p_3)=(0,0)$ & & \qquad \quad \ (b) \ \ $(p_2,p_3)=(1,0)$ \\ \\
\includegraphics[scale=0.32]{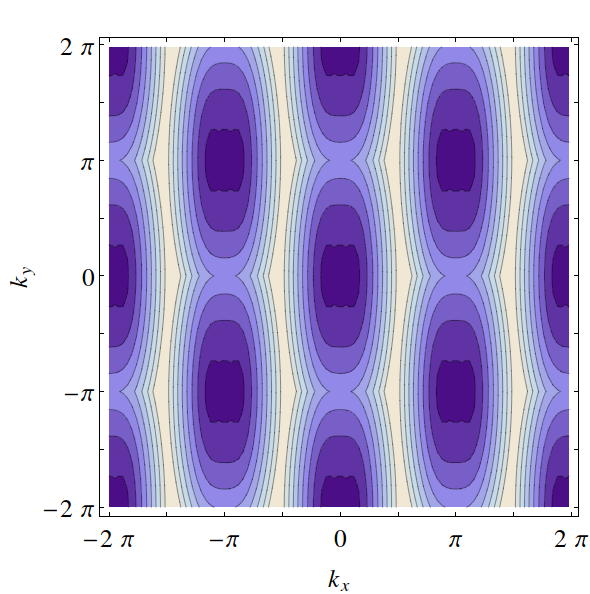} & & \includegraphics[scale=0.32]{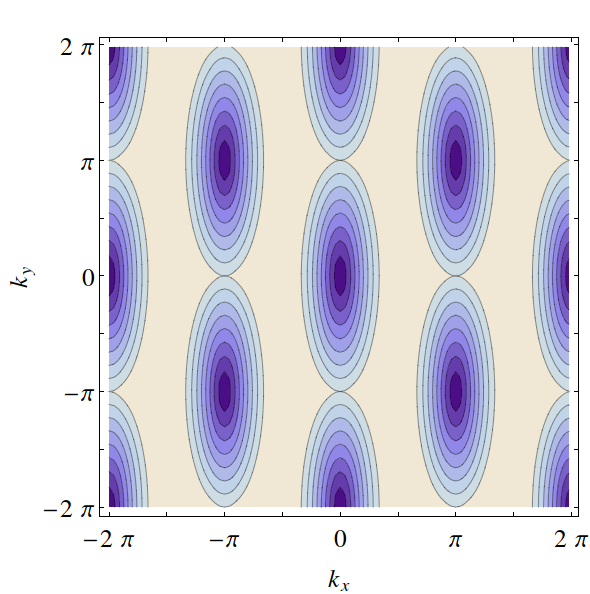} \\
\qquad \quad \ (c) \ \ $(p_2,p_3)=(0,1)$ & & \qquad \quad \ (d) \ \ $(p_2,p_3)=(1,1)$
\end{tabular}
\caption{\label{twospinonspectrumfigure} The dispersion of the lower boundary of two-spinon continuum for four bosonic spin liquid states labelled by $(p_2,p_3)$, at $\kappa$ close to $\kappa_c$. In the plots, $k_x$ and $k_y$ range from $-2 \pi$ to $2 \pi$, while the Brillouin zone is given by $k_x,k_y \in \left[-\pi,\pi\right]$. Darker region indicates lower energy. (a) $(0,0)$ state at $\kappa=0.32$, where $\kappa_c=0.327$. The minimum occurs at $\mathbf{k}=(0,0),(\pm 0.84 \pi,0)$. (b) $(1,0)$ state at $\kappa=0.34$, where $\kappa_c=0.351$. The minimum occurs at $\mathbf{k}=(0,0),(\pm 0.88 \pi,0)$. (c) $(0,1)$ state at $\kappa=0.35$, where $\kappa_c=0.362$. The minimum occurs at $\mathbf{k}=(0,0),(\pm 0.88 \pi, \pm \pi),(\pm \pi, \pm \pi)$. (d) $(p_2,p_3)=(1,1)$ state at $\kappa=0.35$, where $\kappa_c=0.368$. The minimum occurs at $\mathbf{k}=(0,0),(\pm \pi, \pm \pi)$.}
\end{figure*}

To gain further insight about these spin liquid phases, we compute the dispersion $\Omega_\mathbf{k}$ of the lower boundary of two-spinon continuum,
which is given by
\begin{equation} \label{twospinonenergydefinition}
\Omega_\mathbf{k} = \min_\mathbf{q} (\omega_\mathbf{q} + \omega_{\mathbf{k}-\mathbf{q}}) ,
\end{equation}
where $\omega_\mathbf{q}$ is the one-spinon dispersion.
We plot $\Omega_\mathbf{k}$ of each spin liquid state with $\kappa$ close to $\kappa_c$, for $-2 \pi \leq k_x,k_y \leq 2 \pi$, as shown in FIG \ref{twospinonspectrumfigure}. Darker regions indicate lower energy.

Notice that $\Omega_\mathbf{k}$ with the same $p_3$ have similar profiles. For $p_3=0$, the minimum of the two-spinon continuum occurs at $\mathbf{k}=(\pm q,0)$ (as well as at the zone center), which suggests that $(q,0)$ magnetic order would develop beyond $\kappa > \kappa_c$. On the other hand, for $p_3=1$, the minimum occurs at $(\pi, \pi)$ (as well as at the zone center), which indicates that $(\pi,\pi)$ magnetic order would arise beyond $\kappa_c$. Details of the two-spinon continuum and the associated magnetic ordering patterns will be discussed in the next two sections.

Finally, it can be explicitly checked that four spin liquid solutions labelled by $(p_2,p_3)$ are time-reversal invariant by computing the flux\cite{PhysRevB.74.174423,PhysRevB.87.125127} piercing through the length-6 hexagon and three independent length-8 rhombi on the non-symmorphic Kagome lattice. We confirm that these fluxes are always $0$ or $\pi$ in our solution, which is a necessary condition for the time reversal invariance. It can be checked that these solutions are indeed related to $(p_2,p_3,p_{13}=0)$ states in the full classification of the time reversal invariant spin liquid phases. Hence the flux counting is consistent with the PSG analysis.

\subsection{\label{classicalorder} Magnetically Ordered States}

When the spin magnitude or the bosonic density $\kappa = 2S$ reaches its critical value $\kappa_c$, the excitation spectrum becomes gapless and spinons condense at particular wavevectors $\mathbf{k}_c$, where $\omega_{\mathbf{k}_c}=0$, causing a phase transition from a spin liquid to a long range magnetically ordered state. The spinor form of the spinon operator $\Psi_\mathbf{k} = (b_{\mathbf{k} \uparrow}, b_{- \mathbf{k} \downarrow}^\dagger)^\mathrm{T}$ (see \eqref{vectorspinonfieldboson} in Appendix \ref{meanfieldcalculation}), gains a finite expectation value at these $\mathbf{k}$ points, which is proportional to the respective critical eigenvector that becomes soft at $\mathbf{k}_c$. Analyzing these eigenmodes, we can determine the real space ordering patterns.\cite{PhysRevB.45.12377,PhysRevB.74.174423} Below we focus on the two most energetically favorable spin liquid phases, namely $(p_2,p_3) = (1,0)$ and $(1,1)$. It is found that $\mathbf{k}_c=\pm(0.44\pi,0)$ for the state $(1,0)$, while $\mathbf{k}_c=\pm(\pi/2,\pi/2)$ for $(1,1)$.

\begin{figure}
\begin{tabular}{>{\centering\arraybackslash} m{6.5 cm}}
\includegraphics[scale=0.3]{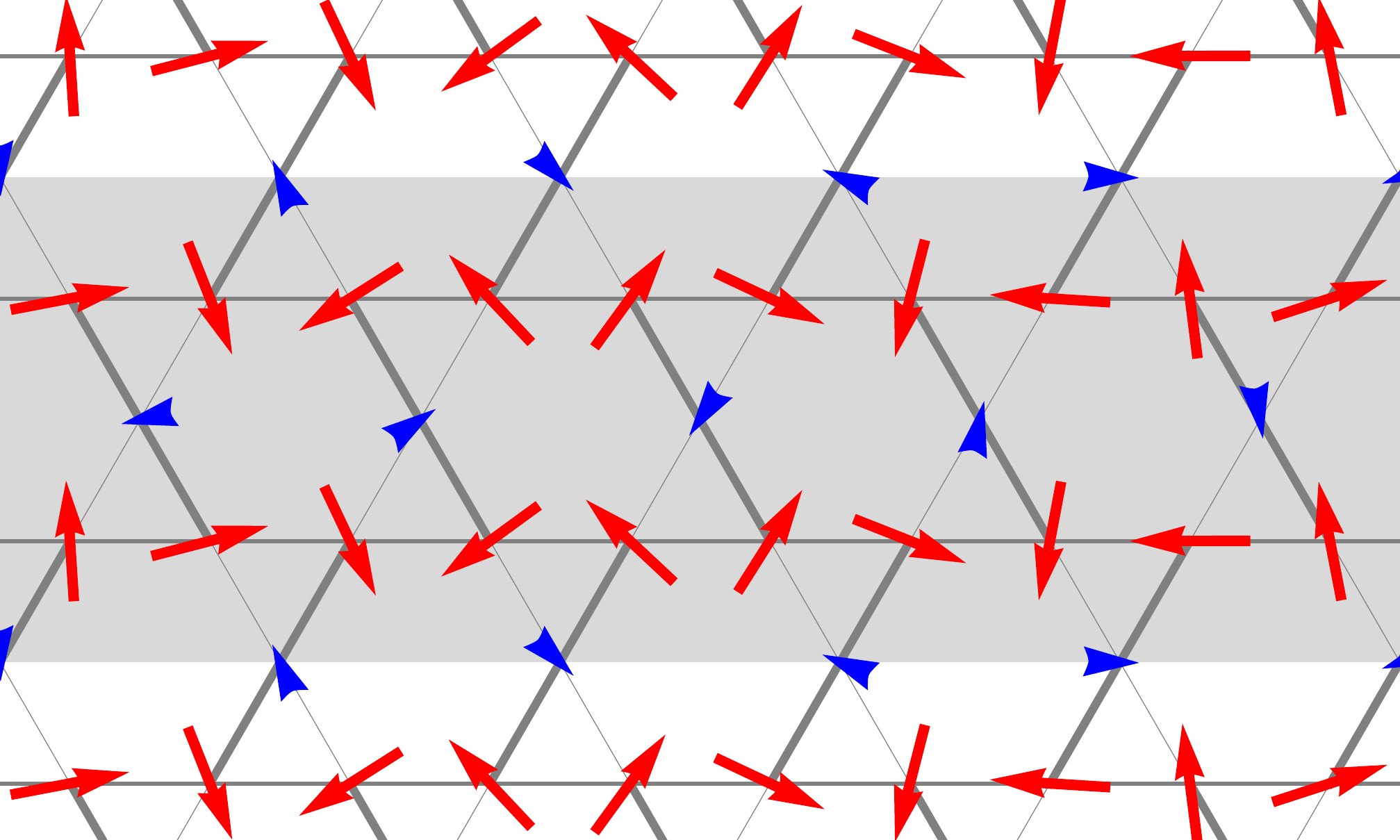} \\ [2 pt]
(a) \ \ $(p_2,p_3)=(1,0)$ \\ [10 pt]
\includegraphics[scale=0.3]{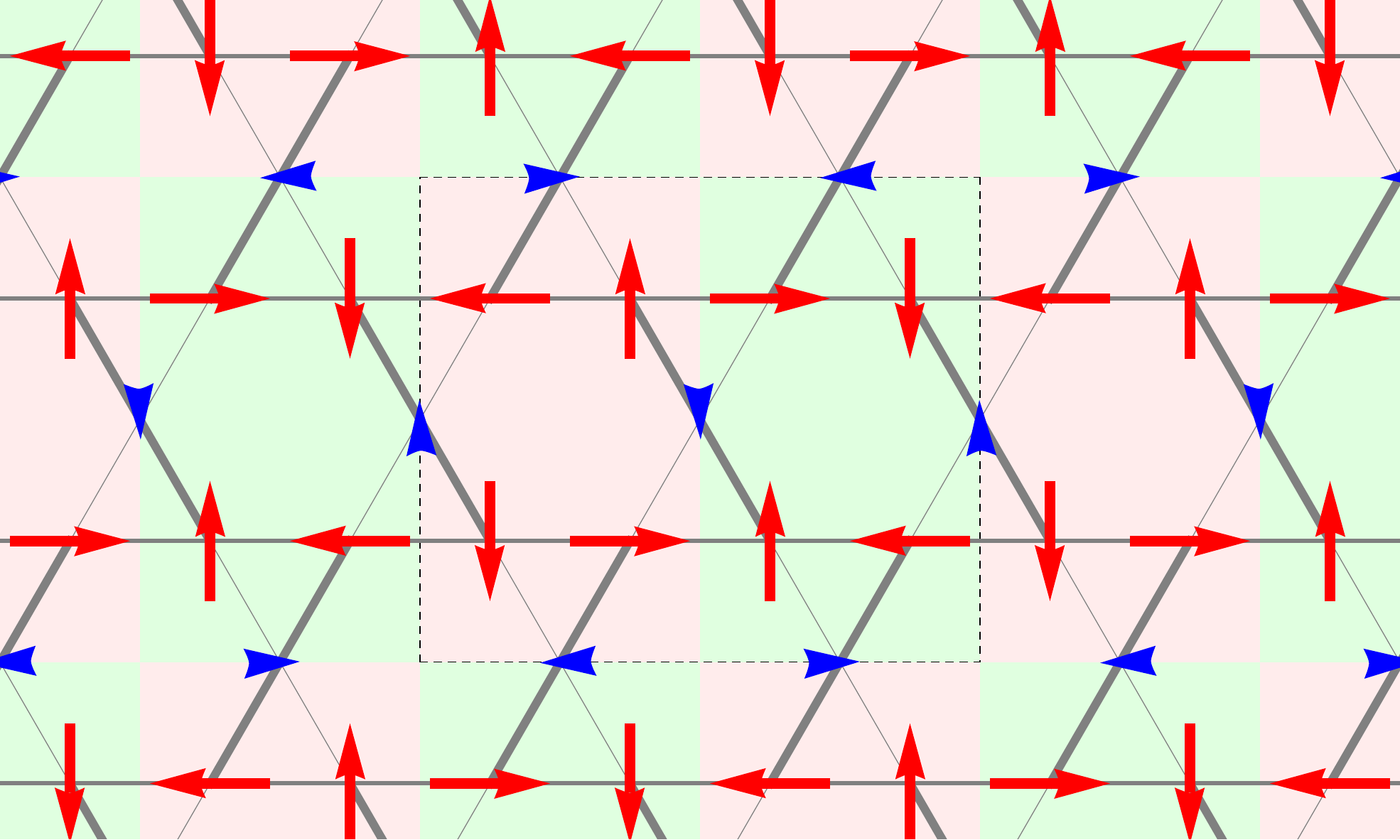} \\ [2 pt]
(b) \ \ $(p_2,p_3)=(1,1)$
\end{tabular}
\caption{\label{spinliquidmagneticorder} Magnetic ordering patterns obtained from the spinon condensation in spin liquid phases. (a) The coplanar incommensurate $(q,0)$ spiral order with $q=0.88 \pi$, obtained from the $(1,0)$ spin liquid state. (b) The coplanar commensurate $(\pi, \pi)$ spin density wave order obtained from the $(1,1)$ spin liquid state.}
\end{figure}

More precisely, we first determine the expectation value of the real space spinor form $x_i \equiv (\langle b_{i \uparrow} \rangle, \langle b_{i \downarrow} \rangle)^\mathrm{T}$ via the Fourier transformation of $\langle \Psi_\mathbf{k} \rangle = (\langle b_{\mathbf{k} \uparrow} \rangle, \langle b_{- \mathbf{k} \downarrow}^\dagger \rangle)^\mathrm{T}$, which is dominated by the contribution at $\mathbf{k} = \mathbf{k}_c$. The real space spin configuration is then obtained from
\begin{equation}
\left\langle \mathbf{S}_i \right\rangle \approx \frac{1}{2} x_i^\dagger \bm{\sigma} x_i ,
\end{equation}
which we plot in FIG. \ref{spinliquidmagneticorder}a and  \ref{spinliquidmagneticorder}b for states $(1,0)$ and $(1,1)$ respectively.

In the case of the (1,0) spin liquid state, a coplanar incommensurate $(q,0)$ spiral order with $q=0.88 \pi$ develops as $\kappa$ reaches the critical value. Here the spins rotate by $0.88 \pi$ under lattice translation $T_x$, while they do not change under $T_y$. The amplitude of spin is the same among the chain (interstitial) sites, but it is larger at the chain sites than at the interstitial sites. On the other hand, a coplanar commensurate $(\pi,\pi)$ spin density wave order develops if one starts from the $(1,1)$ spin liquid state. Here the spins rotate by $\pi$ under lattice translation $T_x$ or $T_y$. The amplitude of spin is the same among the chain (interstitial) sites, but it is larger at the chain sites than at the interstitial sites. Notice that in both configurations, the spins which interact by the dominant antiferromagnetic coupling $J$ are anti-aligned.

\begin{figure}
\includegraphics[scale=0.3]{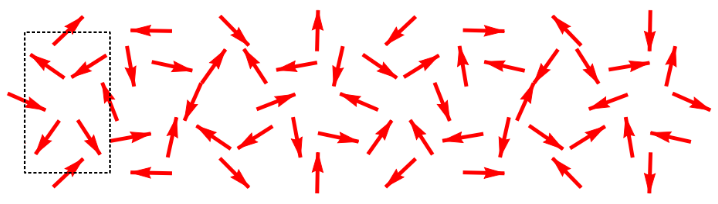}
\caption{\label{classicalmagneticorder} Magnetic ordering pattern for the $(q,0)$ spiral order found in the classical model. The dotted box indicates the unit cell. This state is essentially the same as the $(q,0)$ magnetic order obtained from the (1,0) spin liquid state via the spinon condensation (see FIG. \ref{spinliquidmagneticorder}a).}
\end{figure}

To elucidate the nature of these magnetically ordered phases, we investigate the ground state of the classical model, where
we treat the spins in the Heisenberg model \eqref{heisenberggeneral} as three-component vectors of fixed length, $\mathbf{S}_i=(S^x_i,S^y_i,S^z_i)$ and $\lvert \mathbf{S}_i \rvert = 1$. We use the same exchange interactions given in Section \ref{microscopic}. Simulated annealing is employed to obtain the ground state spin configuration on a lattice with $64 \times 32$ unit cells, which turns out to be a coplanar $(q,0)$ spiral order with $q=3\pi/4$, as shown in FIG. \ref{classicalmagneticorder}.

Notice that the ground state spin configuration in the classical Heisenberg model agrees quite well with that arising from the most energetically favorable spin liquid state $(1,0)$, except the numerical value of $q$ in the ordering wave vector and the uniformity of spin amplitude. Hence it is natural to conclude that the $(q,0)$ spiral order derived from the (1,0) spin liquid state is continuously connected to the classical limit. It is, therefore, natural to call this state a $(q,0)$ spiral order. On the other hand, the spin density wave order arising from the $(1,1)$ spin liquid state has no classical analog. Hence the emergence of this state is a purely quantum effect as this state can only appear as the ground state if the amplitude of spins varies over different sites. This is the reason why we call this state a $(\pi,\pi)$ spin density wave. Since the local moments of Volborthite carry $S=1/2$, this material is not close to the classical limit, which suggests that the $(q,0)$ spiral and $(\pi,\pi)$ spin density wave states may be competing magnetic orders below $1 \, \mathrm{K}$ in Volborthite.

\subsection{\label{enhancedperiodicity} Periodicity Enhancement of Two-Spinon Continuum}
The dispersion of the lower boundary of two-spinon continuum is shown in FIG. \ref{twospinonspectrumfigure} for different spin liquid phases.
It can be seen that the spectra with $p_3=1$ exhibit periodicity enhancement, namely the translation by $\mathbf{k}=(\pm \pi, \pm \pi)$ in momentum space leaves the spectra invariant, which is beyond the periodicity allowed by the lattice translational symmetry.
Such an enhanced periodicity, which leads to the spectral doubling,\cite{PhysRevB.90.121102} is a consequence of the spatial inversion $C_2$ and time reversal $\mathcal{T}$ symmetries. To prove this, we first observe that a physical spin operator, which is a bilinear form of two spinon operators, must transform trivially under the symmetry operations that amount to identity. However, each spinon can transform projectively under the same symmetry operations.\cite{1403.0575} It can gain a phase of $\pm \pi$ which is determined by the solution of algebraic PSG. In particular, the $\mathbb{Z}_2$ variable $p_3$ characterizes how a spinon transforms under the algebraic identities \eqref{algebraicrelation3} and \eqref{algebraicrelation4}, i.e. $\tilde{C}_2 \tilde{T}_x \tilde{C}_2^{-1} \tilde{T}_x = (-1)^{p_3}$ and $\tilde{C}_2 \tilde{T}_y {\tilde{C}_2}^{-1} \tilde{T}_y = (-1)^{p_3}$ (see Appendix \ref{bosonicpsgderive}), where $\tilde{X} \equiv G_X X$.

\begin{figure}
\includegraphics[scale=0.31]{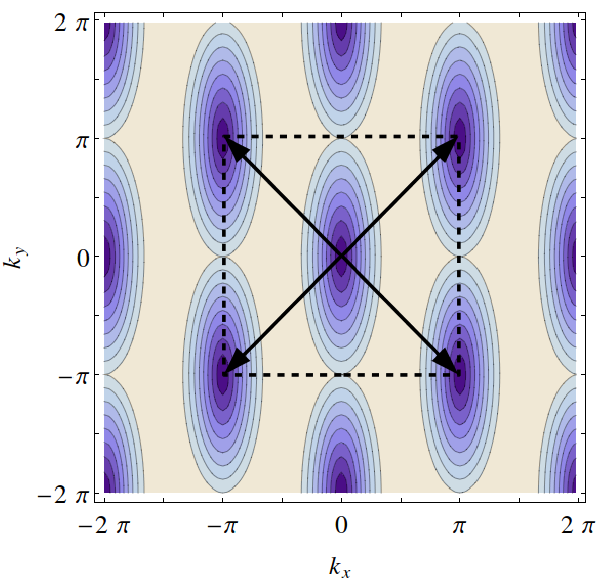}
\caption{\label{enhancedperiodicityfigure} Two-spinon spectrum for $(p_2,p_3)=(1,1)$ spin liquid state, which exhibits periodicity enhancement. Translation by $\mathbf{k}=(\pm \pi, \pm \pi)$ in momentum space leaves the spectra invariant. Dashed line indicates the Brillouin zone.}
\end{figure}

Let us denote a one-spinon momentum eigenstate by $\lvert \mathbf{q} \rangle$, which is necessarily an energy eigenstate with eigenvalue $\omega_\mathbf{q}$. Since momentum is the generator of translation, we get $\tilde{T}_a \lvert \mathbf{q} \rangle = \exp (iq_a) \lvert \mathbf{q} \rangle$ where $a=x,y$. Consider the state $\lvert \mathbf{q}' \rangle = \tilde{\mathcal{T}} \tilde{C}_2 \lvert \mathbf{q} \rangle$, which is degenerate with $\lvert \mathbf{q} \rangle$ although in general $\mathbf{q}' \neq \mathbf{q}$. To see how $\mathbf{q}'$ is related to $\mathbf{q}$, we apply the one-spinon translation operator $\tilde{T}_a$,
\begin{equation} \label{spectraldoublingderive}
\begin{aligned}[b]
\tilde{T}_a \lvert \mathbf{q}' \rangle &= \tilde{T}_a \tilde{\mathcal{T}} \tilde{C}_2 \lvert \mathbf{q} \rangle \\
&= \tilde{\mathcal{T}} \tilde{T}_a \tilde{C}_2 \lvert \mathbf{q} \rangle \\
&= \tilde{\mathcal{T}} (-1)^{p_3} \tilde{C}_2 {\tilde{T}_a}^{-1} \lvert \mathbf{q} \rangle \\
&= e^{i p_3 \pi} \tilde{\mathcal{T}} \tilde{C}_2 e^{-iq_a} \lvert \mathbf{q} \rangle \\
&= e^{i(q_a + p_3 \pi)} \lvert \mathbf{q}' \rangle ,
\end{aligned}
\end{equation}
where we have used the fact that translation commutes with time reversal at one-spinon level, i.e. $\tilde{T}_a \tilde{\mathcal{T}} = \tilde{\mathcal{T}} \tilde{T}_a$, from our solution of the algebraic PSG (see Appendix \ref{bosonicpsgderive}). \eqref{spectraldoublingderive} tells us that $\mathbf{q}'=\mathbf{q} + p_3 (\pi,\pi)$ while $\omega_{\mathbf{q}'} = \omega_\mathbf{q}$.
Since the two-spinon energy takes the form $\Omega_{\mathbf{p}+\mathbf{q}} = \omega_\mathbf{p} + \omega_{\mathbf{q}}$, we have
\begin{equation}
\begin{aligned}[b]
\Omega_{\mathbf{p}+\mathbf{q}+p_3(\pi,\pi)} &= \omega_\mathbf{p} + \omega_{\mathbf{q}+p_3(\pi,\pi)} \\
&= \omega_\mathbf{p} + \omega_{\mathbf{q}} \\
&= \Omega_{\mathbf{p}+\mathbf{q}} .
\end{aligned}
\end{equation}
When $p_3=0$, this equation is trivially satisfied. When $p_3=1$, we have $\Omega_{\mathbf{k}+(\pi,\pi)} = \Omega_\mathbf{k}$, which is the periodicity enhancement we observe in, for example, FIG. \ref{enhancedperiodicityfigure}. We have thus shown that the enhanced periodicity arises from the symmetry that combines spatial inversion $C_2$ and time reversal $\mathcal{T}$. This class of enhanced periodicity has been discussed in Ref.~\onlinecite{PhysRevB.90.121102} and explained through the language of symmetry fractionalization,\cite{PhysRevB.87.104406} where the symmetry action on a composite physical operator/state (e.g. spin) can be represented by the product of individual symmetry action on each constituent spinon operator/state. Hence our result can also be interpreted as a consequence of symmetry fractionalization.

\section{\label{summary} Discussion}

In this work, we investigate possible quantum spin liquid and magnetically ordered phases in a two-dimensional non-symmorphic lattice, motivated by the experiments on Volborthite, $\mathrm{Cu_3 V_2 O_7 (OH)_2 \cdot 2 H_2 O}$, and the DFT computations. In Volborthite, a magnetic ordering \cite{PhysRevLett.103.077207,PhysRevLett.114.227202,1602.04028} occurs below $1 \, \mathrm{K}$ while a recent thermal conductivity measurement \cite{1608.00444} found some signatures of possible spin liquid behavior right above $1 \, \mathrm{K}$. This may suggest that the paramagnetic state above $1 \, \mathrm{K}$ may be proximate to a putative quantum spin liquid that is intimately related to the magnetic order below $1 \, \mathrm{K}$. The nature of the magnetic order below $1 \, \mathrm{K}$, however, is not fully understood. Earlier nuclear magnetic resonance (NMR) measurements suggest that the magnetic order could be a helical order or a spin density wave.

To address these issues, we use the Schwinger boson mean field theory \cite{PhysRevB.45.12377,PhysRevB.74.174423,PhysRevB.87.125127} and PSG \cite{PhysRevB.65.165113} to understand the connection between possible magnetic order and quantum spin liquid phases on equal footing, in the spin model \cite{PhysRevLett.117.037206} derived from the DFT computation. The DFT-derived model suggests that the underlying lattice has non-symmorphic symmetry, described by $p2gg$ planar space group. We first show that mean field spin liquid states in two-dimensional non-symmorphic lattices do not enlarge the lattice unit cell because of the glide symmetry, the combination of refection and a fractional translation. We analyze the resulting mean field ground states and find that there are two competing quantum spin liquid states that lead to a $(q,0)$ spiral magnetic order and a $(\pi,\pi)$ spin density wave, respectively, upon the spinon condensation. This suggests that the $(q,0)$ spiral order and $(\pi,\pi)$ spin density wave would be competing magnetically ordered state below $1 \, \mathrm{K}$, and two spin liquid phases mentioned above are candidate paramagnetic states above $1 \, \mathrm{K}$. Notice that both of the $(q,0)$ spiral order and $(\pi,\pi)$ spin density wave would be consistent with earlier NMR experiments. It is also found that the spin liquid phase related to the $(\pi,\pi)$ spin density wave shows periodicity enhancement of the two-spinon continuum, a signature of the so-called symmetry fractionalization.\cite{PhysRevB.87.104406,PhysRevB.90.121102} Hence a future neutron scattering experiment may be able to find such signatures in case that this spin liquid state is relevant to the paramagnetic state above $1 \, \mathrm{K}$.
 
Finally, the spin liquid states with bosonic spinons in two dimensions are necessarily gapped phases as a magnetic ordering will occur if the spin-carrying bosons become gapless. While a small excitation gap at finite temperature in the spin liquid phases with bosonic spinons may explain the large thermal conductivity discovered in Volborthite, it will also be useful to consider fermionic versions of these spin liquid phases, which naturally allows gapless spin liquid states and a large thermal conductivity. This would be an excellent topic of future study.

\begin{acknowledgments}
We thank Sopheak Sorn for initial collaboration and Robert Schaffer for useful discussion.
This work was supported by the NSERC of Canada and the Center for Quantum Materials at the University of Toronto.
T. M. was supported by Advanced Leading Graduate Course for Photon Science (ALPS).
Some of the computations were performed on the GPC supercomputer at the SciNet HPC Consortium.\cite{1742-6596-256-1-012026} SciNet is funded by: the Canada Foundation for Innovation under the auspices of Compute Canada; the Government of Ontario; Ontario Research Fund - Research Excellence; and the University of Toronto.
\end{acknowledgments}

\appendix
\section{\label{spacegroupalgebra} Space Group and Algebraic Identities of Non-Symmorphic Kagome Lattice}
Below we list the action of space group elements, $T_x$, $T_y$, $C_2$ and $h$, on coordinate $(x,y,s)$.
\begingroup
\allowdisplaybreaks
\begin{align*}
T_x: \: & (x,y,s) \longrightarrow \left(x+1,y,s\right) . \\ \\
T_y: \: & (x,y,s) \longrightarrow \left(x,y+1,s\right) . \\ \\
C_2: \: & (x,y,1) \longrightarrow (-x,-y,5) , \\
& (x,y,2) \longrightarrow (-x,-y,4) , \\
& (x,y,4) \longrightarrow (-x,-y,2) , \\
& (x,y,5) \longrightarrow (-x,-y,1) , \\
& (x,y,3) \longrightarrow (-x+1,-y,3) , \\
& (x,y,6) \longrightarrow (-x,-y+1,6) . \\ \\
h: \: & (x,y,1) \longrightarrow (x,-y-1,2) , \\
& (x,y,2) \longrightarrow (x+1,-y-1,1) , \\
& (x,y,3) \longrightarrow (x,-y,6) , \\
& (x,y,6) \longrightarrow (x+1,-y,3) , \\
& (x,y,4) \longrightarrow (x,-y,5) , \\
& (x,y,5) \longrightarrow (x+1,-y,4) .
\end{align*}
\endgroup
The algebraic relations among $T_x$, $T_y$, $C_2$ and $h$ are
\begingroup
\allowdisplaybreaks
\begin{align}
T_x^{-1} T_y^{-1} T_x T_y &= I , \label{algebraicrelation1} \\
C_2^2 &= I , \label{algebraicrelation2} \\
C_2 T_x C_2^{-1} T_x &= I , \label{algebraicrelation3} \\
C_2 T_y C_2^{-1} T_y &= I , \label{algebraicrelation4} \\
T_x^{-1} h^2 &= I , \label{algebraicrelation5} \\
h^{-1} T_x^{-1} h T_x &= I , \label{algebraicrelation6} \\
h^{-1} T_y h T_y &= I , \label{algebraicrelation7} \\
T_x T_y h^{-1} C_2 h C_2 &= I . \label{algebraicrelation8}
\end{align}
\endgroup
If time reversal symmetry is considered, we further have
\begin{align}
\mathcal{T}^2 &= I , \label{algebraicrelation9} \\
X^{-1} \mathcal{T}^{-1} X \mathcal{T} &= I,\, X \in \lbrace T_x, T_y, C_2, h \rbrace . \label{algebraicrelation10}
\end{align}
\eqref{algebraicrelation1} $-$ \eqref{algebraicrelation10} constrain the possible form of gauge transformation $G_X$ associated with symmetry operation $X=T_x,T_y,C_2,h,\mathcal{T}$ such that $G_XX \in \mathrm{PSG}$. Note that \eqref{algebraicrelation5} and \eqref{algebraicrelation6} are not independent of each other, but we are going to use both forms to find the solution to algebraic PSG in Appendix \ref{bosonicpsgderive}.

\section{\label{bosonicpsgderive} Solution to Bosonic PSG}
Here we determine the form of gauge transformations $G_X, X=T_x,T_y,C_2,h,\mathcal{T}$ that are consistent with the algebraic constraints imposed by the identities \eqref{algebraicrelation1} $-$ \eqref{algebraicrelation10}. Relevant discussion can be found in Section \ref{algebraicpsgboson}. We represent the gauge transformation by its phase,
\begin{equation}
G_X(x,y,s) = e^{i\phi_X (x,y,s)} .
\end{equation}

First and foremost, we follow the usual convention \cite{PhysRevB.74.174423} of fixing
\begin{equation*}
\phi_{T_x}(x,y,s) = 0 .
\end{equation*}
The algebraic identity \eqref{algebraicrelation1} leads to
\begin{equation*}
\begin{aligned}[b]
& \phi_{T_y}(x-1,y,s) + \phi_{T_x}(x,y,s) \\
& \qquad - \phi_{T_y}(x,y,s) - \phi_{T_x}(x,y-1,s) = p_1 \pi ,
\end{aligned}
\end{equation*}
where $p_n = 0$ or $1\,\mathrm{mod}\,2$. Simplifiying yields
\begin{equation*}
\phi_{T_y}(x,y,s) = \phi_{T_y}(0,y,s) + p_1 \pi x .
\end{equation*}
Using gauge freedom, we can fix $\phi_{T_y}(0,y,s)=0$. Therefore,
\begin{equation*}
\phi_{T_y}(x,y,s) = p_1 \pi x .
\end{equation*}

Before we proceed to the calculation of rotation and glide, we can eliminate additional parameters in \eqref{algebraicrelation5} and \eqref{algebraicrelation8} by a global $\mathbb{Z}_2$ gauge. Schematically, \eqref{algebraicrelation8} gives
\begin{equation} \label{p8elimination}
\phi_{T_x} + \phi_{T_y} + \ldots = p_8 \pi ,
\end{equation}
where $\ldots$ denotes some undetermined phases of $C_2$ and $h$. The trick is to add a global phase of $p_8 \pi$ to $\phi_{T_y}(x,y,s)$, which does no harm because the IGG is $\mathbb{Z}_2$. Now $p_8 \pi$ appears on both sides of \eqref{p8elimination}, so they cancel out each other. Since all other identities involve an even number of $T_y$, without loss of generality we can set $p_8 = 0\,\mathrm{mod}\,2$. Schematically, \eqref{algebraicrelation5} gives
\begin{equation} \label{p5elimination}
- \phi_{T_x} + \ldots = p_5 \pi ,
\end{equation}
where $\ldots$ denotes some undetermined phases of $h$. Similar to previous case, we add a global phase of $p_5 \pi$ to $\phi_{T_x}(x,y,s)$. \eqref{p5elimination} then forces $p_5 = 0\,\mathrm{mod}\,2$.
\\ \\
From \eqref{algebraicrelation3} and \eqref{algebraicrelation4}, we have
\begin{equation*}
\phi_{C_2} (x,y,s) = \phi_{C_2} (0,0,s) + p_3 \pi x + \left( p_4 + \delta_{s,3} p_1 \right) \pi y .
\end{equation*}
From \eqref{algebraicrelation2}, we have
\begin{equation*}
\begin{aligned}
\phi_{C_2}(0,0,5) + \phi_{C_2}(0,0,1) &= p_2 \pi , \\
\phi_{C_2}(0,0,4) + \phi_{C_2}(0,0,2) &= p_2 \pi , \\
\phi_{C_2}(0,0,3) + \phi_{C_2}(0,0,3) &= ( p_2 + p_3 ) \pi , \\
\phi_{C_2}(0,0,6) + \phi_{C_2}(0,0,6) &= ( p_2 + p_4 ) \pi .
\end{aligned}
\end{equation*}
Consider the sublattice dependent gauge transformation defined by
\begin{equation*}
\begin{aligned}
G_1: \: & \phi_1(x,y,s=1,2) = - \phi_1 , \\
& \phi_1(x,y,s=4,5) = \phi_1 , \\
& \phi_1(x,y,s=3,6) = 0 .
\end{aligned}
\end{equation*}
$\phi_{T_x}$ and $\phi_{T_y}$ are invariant under $G_1$, while $\phi_{C_2}$ is modified by
\begin{equation*}
\begin{aligned}
\phi_{C_2} (x,y,s=1,2) &\longrightarrow \phi_{C_2} (x,y,s=1,2) - 2 \phi_1 , \\
\phi_{C_2} (x,y,s=4,5) &\longrightarrow \phi_{C_2} (x,y,s=4,5) + 2 \phi_1 .
\end{aligned}
\end{equation*}
Therefore, by an appropriate choice of $\phi_1$, we can fix
\begin{equation*}
\begin{aligned}
\phi_{C_2} (0,0,s=1,2) &= 0 , \\
\phi_{C_2} (0,0,s=4,5) &= p_2 \pi .
\end{aligned}
\end{equation*}
Furthermore, with the gauge transformations
\begin{equation*}
\begin{aligned}
G_2: \: & \phi_2(x,y,s) = \pi x , \\
G_3: \: & \phi_3(x,y,s) = \pi y ,
\end{aligned}
\end{equation*}
we can fix
\begin{equation*}
\begin{aligned}
\phi_{C_2} (0,0,3) &= \frac{p_2 + p_3}{2} \pi , \\
\phi_{C_2} (0,0,6) &= \frac{p_2 + p_4}{2} \pi .
\end{aligned}
\end{equation*}
From \eqref{algebraicrelation6} and \eqref{algebraicrelation7}, we have
\begin{equation*}
\begin{aligned}
\phi_h(x,y,s) = & \phi_h\left(0,0,s\right) + p_6 \pi x \\
& + \left( p_7 + p_1 \left( \delta_{s,1} + \delta_{s,3} + \delta_{s,4} \right) \right) \pi y ,
\end{aligned}
\end{equation*}
From \eqref{algebraicrelation5}, starting from $(x,y,1)$, we have
\begin{equation*}
\phi_h(0,0,2) + \phi_h(0,0,1) + p_6 \pi - p_7 \pi + p_1 \pi y = 0 .
\end{equation*}
Since R.H.S. has no coordinate dependence, this forces $p_1 = 0 \,\mathrm{mod}\, 2$. From \eqref{algebraicrelation5}, starting from $(x,y,4)$,
\begin{equation*}
\phi_h(0,0,5) + \phi_h(0,0,4) + p_6 \pi = 0 .
\end{equation*}
From \eqref{algebraicrelation5}, starting from $(x,y,5)$, we have
\begin{equation*}
\phi_h(0,0,4) + \phi_h(0,0,5) = 0 .
\end{equation*}
The two equations above implies that $p_6 = 0 \,\mathrm{mod}\, 2$. From \eqref{algebraicrelation8}, starting from $(x,y,3)$, we have
\begin{equation*}
\phi_{C_2}(0,0,3) + \phi_{C_2}(0,0,6) + p_4 \pi + p_7 \pi = 0 .
\end{equation*}
From \eqref{algebraicrelation8}, starting from $(x,y,6)$, we have
\begin{equation*}
\phi_{C_2}(0,0,6) + \phi_{C_2}(0,0,3) = 0 .
\end{equation*}
The two condition above implies that $p_7 = p_4 \,\mathrm{mod}\, 2$. Multiplying the last equation by 2 gives
\begin{equation*}
p_3 \pi + p_4 \pi = 0 .
\end{equation*}
Therefore $p_4 = p_3 \,\mathrm{mod}\, 2$. To eliminate $p_4$ in favor of $p_3$ in $\phi_{C_2}(0,0,6)$, we have to treat division by 2 carefully since there may be an additional phase of $\pi$, 
\begin{equation*}
\phi_{C_2}(0,0,6)=\frac{p_2+p_3}{2} \pi + q \pi ,
\end{equation*}
where $q = 0$ or $1 \,\mathrm{mod}\, 2$. $q$ can be expressed in terms of $p_2$ and $p_3$ by
\begin{equation*}
\begin{aligned}
0 &= \phi_{C_2}(0,0,3) + \phi_{C_2}(0,0,6) \\
&= p_2 \pi + p_3 \pi + q \pi .
\end{aligned}
\end{equation*}
or $q=\left(p_2+p_3\right) \,\mathrm{mod}\, 2$. Therefore,
\begin{equation*}
\phi_{C_2}(0,0,6)=\frac{3(p_2+p_3)}{2} \pi .
\end{equation*}
From \eqref{algebraicrelation8}, starting from $(x,y,1)$,
\begin{equation*}
\phi_h(0,0,4) - \phi_h(0,0,2) + p_2 \pi - p_3 \pi  = 0 ,
\end{equation*}
From \eqref{algebraicrelation8}, starting from $(x,y,2)$,
\begin{equation*}
\phi_h(0,0,5) - \phi_h(0,0,1) + p_2 \pi = 0 .
\end{equation*}
Evaluating \eqref{algebraicrelation8} starting from $(x,y,s=4,5)$ generates the same two equations above.
\\ \\
From \eqref{algebraicrelation5}, starting from $(x,y,3)$,
\begin{equation*}
\phi_h(0,0,6) + \phi_h(0,0,3) = 0 .
\end{equation*}
Now, we should determine the form of $\phi_h(0,0,s)$. Using the  sublattice dependent gauge transformation defined by
\begin{equation*}
\begin{aligned}
G_4: \: & \phi_4\left(x,y,s=3\right) = \phi_4 , \\
& \phi_4\left(x,y,s=6\right) = - \phi_4 , \\
& \phi_4\left(x,y,s=1,2,4,5\right) = 0 .
\end{aligned}
\end{equation*}
we first fix
\begin{equation*}
\phi_h(0,0,3) = 0 = \phi_h(0,0,6) .
\end{equation*}
We are then left with the system of linear equations from previous calculations
\begin{equation*}
\left\lbrace
\begin{aligned}
\phi_h(0,0,1) + \phi_h(0,0,4) &= p_2 \pi , \\
\phi_h(0,0,1) + \phi_h(0,0,2) &= p_3 \pi , \\
\phi_h(0,0,4) + \phi_h(0,0,5) &= 0 .
\end{aligned} \right.
\end{equation*}
Using a similar sublattice dependent gauge transformation, we can fix
\begin{equation*}
\phi_h(0,0,4) = 0 = \phi_h(0,0,5) ,
\end{equation*}
which then implies
\begin{equation*}
\begin{aligned}
\phi_h(0,0,1) &= p_2 \pi , \\
\phi_h(0,0,2) &= (p_2 + p_3) \pi .
\end{aligned}
\end{equation*}
The final solution to algebraic PSG involving only spatial symmetries is given by
\begin{align}
\phi_{T_x}(x,y,s) &= 0 , \label{translationxpsgbosonappendix} \\
\phi_{T_y}(x,y,s) &= 0 , \label{translationypsgbosonappendix} \\
\phi_{C_2}(x,y,s) &= \phi_{C_2}\left(0,0,s\right) + p_3 \pi \left( x + y \right) , \label{rotationpsgbosonappendix} \\
\phi_h(x,y,s) &= \phi_h\left(0,0,s\right) + p_3 \pi y , \label{glidepsgbosonappendix}
\end{align}
with
\begin{equation*}
\begin{aligned}
\phi_{C_2} (0,0,s=1,2) &= 0 , \\
\phi_{C_2} (0,0,s=4,5) &= p_2 \pi , \\
\phi_{C_2} (0,0,s=3) &= \frac{p_2+p_3}{2} \pi , \\
\phi_{C_2} (0,0,s=6) &= \frac{3(p_2+p_3)}{2} \pi , \\
\phi_h (0,0,s=1) &= p_2 \pi , \\
\phi_h (0,0,s=2) &= (p_2+p_3) \pi , \\
\phi_h (0,0,s=3,4,5,6) &= 0 ,
\end{aligned}
\end{equation*}
which is described by only two $\mathbb{Z}_2$ variables $p_2$ and $p_3$.

\subsection{\label{timereversalpsgboson} Time Reversal Symmetry in Bosonic PSG}
Conventionally, time reversal symmetry is enforced by flux argument in Schwinger boson mean field theory, unlike the fermionic approach \cite{PhysRevB.65.165113,PhysRevB.83.224413,1610.06191} which involves time reversal symmetry directly in PSG calculation together with the space group. Since Lu \textit{et al} proposed a scheme of unifying bosonic and fermionic theories of spin liquid through vison PSG,\cite{1403.0575} there have been several attempts \cite{1603.03041,1605.05322,PhysRevB.94.035160} to treat time reversal symmetry in bosonic PSG on equal footing with fermionic PSG. Below we clarify such an approach and solve the bosonic PSG from algebraic identities \eqref{algebraicrelation9} and \eqref{algebraicrelation10} involving time reversal $\mathcal{T}$, so that we can establish connection between bosonic and fermionic spin liquid states in the future.

In analogy to fermionic PSG, we can define the action of $G_X X \in \mathrm{PSG}$ on the pairing ansatz $A_{ij}$ for a space group element $X$ as
\begin{equation} \label{pairingspacegrouppsgactionboson}
G_X X: A_{ij} \longrightarrow e^{-i\phi_X(i)} A_{X^{-1}(i) X^{-1}(j)} e^{-i\phi_X(j)} ,
\end{equation}
while the hopping ansatz $B_{ij}$ follows a similar transformation rule. Note that \eqref{pairingspacegrouppsgactionboson} is consistent with \eqref{pairingbosonpsgrelation} and the derivation of algebraic PSG above. As argued in Ref.~\onlinecite{PhysRevB.65.165113}, time reversal operator complex conjugates the ansatz
\begin{equation} \label{timereversalcomplexconjugate}
\mathcal{T}: A_{ij} \longrightarrow A_{ij}^* .
\end{equation}
In fermionic PSG, when the ansatz is complex conjugated by time reversal, we have the freedom to perform $SU(2)$ gauge transformation $i \tau^2$ to trade the complex conjugation for a minus sign.\cite{PhysRevB.65.165113} However, such an $SU(2)$ gauge redundancy is not present in the bosonic Hamiltonian \eqref{heisenbergmeanfieldboson}, so the complex conjugation \eqref{timereversalcomplexconjugate} cannot be removed. If the system has time reversal symmetry, then $A_{ij}$ and $A_{ij}^*$ describe the same physical state, and thus by the argument in Section \ref{bosonicpsganalysis} they must be equal up to a $U(1)$ gauge transformation, say
\begin{equation} \label{timereversalgaugedefinition}
A_{ij}^* = e^{i \phi_\mathcal{T}(i)} A_{ij} e^{i \phi_\mathcal{T}(j)} .
\end{equation}
Recall that PSG is defined as the group of compound operators $G_X X$ that leave the mean field ansatz invariant. For time reversal, the PSG element $G_\mathcal{T} \mathcal{T}$ is constructed such that, acting $\mathcal{T}$ first complex conjugates the ansatz, acting $G_\mathcal{T}$ next brings the complex conjugated ansatz back to the original one. Using \eqref{timereversalgaugedefinition},
\begin{equation*}
A_{ij} \overset{\mathcal{T}}{\longrightarrow} A_{ij}^* \overset{G_\mathcal{T}}{\longrightarrow} e^{-i \phi_\mathcal{T}(i)} A_{ij}^* e^{-i \phi_\mathcal{T}(j)} \equiv A_{ij} ,
\end{equation*}
or simply (c.f. \eqref{pairingspacegrouppsgactionboson})
\begin{equation}
G_\mathcal{T} \mathcal{T}: A_{ij} \longrightarrow e^{-i \phi_\mathcal{T}(i)} A_{ij}^* e^{-i \phi_\mathcal{T}(j)} .
\end{equation}
The action of $\left(G_\mathcal{T} \mathcal{T}\right)^{-1}$ is naturally first multiply the inverse phase factor to the ansatz and then complex conjugate everything. The net effect of $\left(G_\mathcal{T} \mathcal{T}\right)^{-1}$ is thus same as $G_\mathcal{T} \mathcal{T}$,
\begin{equation}
\left( G_\mathcal{T} \mathcal{T} \right)^{-1}: A_{ij} \longrightarrow e^{-i \phi_\mathcal{T}(i)} A_{ij}^* e^{-i \phi_\mathcal{T}(j)} .
\end{equation}

Suppose that $G_X X \in \mathrm{PSG}$ for a space group element $X$. Applying a gauge transformation on the mean field ansatz, $A_{ij} \longrightarrow G A_{ij}$, $G_X$ changes as \eqref{gaugetransformpsgspaceboson} such that $G_X X$ is still a PSG element. However, since the time reversal operator complex conjugates the ansatz, it is easy to see that for the same gauge transformation above, $G_\mathcal{T}$ changes by
\begin{equation} \label{gaugetransformpsgtimeboson}
\phi_\mathcal{T} (\mathbf{r}) \longrightarrow \phi_\mathcal{T} (\mathbf{r}) + 2 \phi (\mathbf{r}) ,
\end{equation}
such that $G_\mathcal{T} \mathcal{T}$ is still a PSG element.

The time reversal satisfies the algebraic identities \eqref{algebraicrelation9} and \eqref{algebraicrelation10}, which constrain the form of $G_\mathcal{T}$ by
\begin{equation}
\left(G_X X\right)^{-1} \left(G_\mathcal{T} \mathcal{T}\right)^{-1} \left(G_X X\right) \left(G_\mathcal{T} \mathcal{T}\right) = e^{i p_{X \mathcal{T}} \pi} \label{xtconstraintboson}
\end{equation}
and
\begin{equation}
\left(G_\mathcal{T} \mathcal{T}\right)^2 = e^{i p_\mathcal{T} \pi} \label{tsquareconstraintboson}
\end{equation}
respectively, where $X=T_x,T_y,C_2,h$ and $p_{X \mathcal{T}},p_\mathcal{T} \in \lbrace 0, 1 \rbrace$. Applying \eqref{xtconstraintboson} to the mean field ansatz step by step,
\begin{widetext}
\begin{equation}
\begin{aligned}[b]
A_{ij} \overset{G_\mathcal{T} \mathcal{T}}{\longrightarrow} & e^{-i \phi_\mathcal{T}(i)} A_{ij}^* e^{-i \phi_\mathcal{T}(j)} \\
\overset{G_X X}{\longrightarrow} & e^{-i \phi_X(i)} e^{-i \phi_\mathcal{T}(X^{-1}(i))} A_{X^{-1}(i)X^{-1}(j)}^* e^{-i \phi_\mathcal{T}(X^{-1}(j))} e^{-i \phi_X(j)} \\
\overset{\left(G_\mathcal{T} \mathcal{T}\right)^{-1}}{\longrightarrow} & e^{-i \phi_\mathcal{T}(i)} e^{i \phi_X(i)} e^{i \phi_\mathcal{T}(X^{-1}(i))} A_{X^{-1}(i)X^{-1}(j)} e^{i \phi_\mathcal{T}(X^{-1}(j))} e^{i \phi_X(j)} e^{-i \phi_\mathcal{T}(j)} \\
\overset{\left(G_X X\right)^{-1}}{\longrightarrow} & e^{i \phi_X(X(i))} e^{-i \phi_\mathcal{T}(X(i))} e^{i \phi_X(X(i))} e^{i \phi_\mathcal{T}(i)} A_{ij} e^{i \phi_\mathcal{T}(j)} e^{i \phi_X(X(j))} e^{-i \phi_\mathcal{T}(X(j))} e^{i \phi_X(X(j))} .
\end{aligned}
\end{equation}
\end{widetext}
To obtain the next line, we have used the fact that the previous line transforms as $A_{ij}$ by the definition of PSG, regardless of how complicated the expression is. This leads to the consistent condition
\begin{equation} \label{xtconstraintbosonphase}
- \phi_\mathcal{T}(X(\mathbf{r})) + \phi_\mathcal{T}(\mathbf{r}) + 2 \phi_X(X(\mathbf{r})) = p_{X \mathcal{T}} \pi .
\end{equation}
On the other hand, applying \eqref{tsquareconstraintboson} to the mean field ansatz yields
\begin{equation}
\begin{aligned}[b]
A_{ij} & \overset{G_\mathcal{T} \mathcal{T}}{\longrightarrow} e^{-i \phi_\mathcal{T}(i)} A_{ij}^* e^{-i \phi_\mathcal{T}(j)} \\
& \overset{G_\mathcal{T} \mathcal{T}}{\longrightarrow} e^{-i \phi_\mathcal{T}(i)} e^{i \phi_\mathcal{T}(i)} A_{ij} e^{i \phi_\mathcal{T}(j)} e^{-i \phi_\mathcal{T}(j)} \\
& = e^{i \pi} e^{-i \phi_\mathcal{T}(i)} e^{i \phi_\mathcal{T}(i)} A_{ij} e^{i \phi_\mathcal{T}(j)} e^{-i \phi_\mathcal{T}(j)} e^{i \pi} .
\end{aligned}
\end{equation}
where we have manually inserted a factor of $-1$ on both sides of $A_{ij}$ in the last line, because $\mathcal{T}^2$ acting on a $S=1/2$ object should produce a minus sign. This leads to
\begin{equation}
\pi - \phi_\mathcal{T}(\mathbf{r}) + \phi_\mathcal{T}(\mathbf{r}) = p_\mathcal{T} \pi
\end{equation}
or $p_\mathcal{T} = 1$.

With the gauge $G_X$ associated with $X=T_x,T_y,C_2,h$ fixed as \eqref{translationxpsgbosonappendix} $-$ \eqref{glidepsgbosonappendix}, we now use \eqref{xtconstraintbosonphase} to solve for $G_\mathcal{T}$. $X=T_x,T_y$ gives
\begin{equation*}
\phi_\mathcal{T} \left(x,y,s\right) = p_{10} \pi x + p_{11} \pi y + \phi_\mathcal{T} \left(0,0,s\right) .
\end{equation*}
$X=C_2$ gives
\begingroup
\allowdisplaybreaks
\begin{align*}
- \phi_\mathcal{T} \left(0,0,1\right) + \phi_\mathcal{T} \left(0,0,5\right) &= p_{12} \pi , \\
- \phi_\mathcal{T} \left(0,0,2\right) + \phi_\mathcal{T} \left(0,0,4\right) &= p_{12} \pi , \\
- p_{10} \pi + \left(p_2 + p_3\right) \pi &= p_{12} \pi , \\
- p_{11} \pi + 3\left(p_2 + p_3\right) \pi &= p_{12} \pi ,
\end{align*}
\endgroup
which implies $p_{11} = p_{10} \, \mathrm{mod} \, 2$ and $p_{12}=\left(p_2+p_3+p_{10}\right) \, \mathrm{mod} \, 2$. Finally, $X=h$ gives
\begingroup
\allowdisplaybreaks
\begin{align*}
p_{10} \pi - \phi_\mathcal{T} \left(0,0,2\right) + \phi_\mathcal{T} \left(0,0,1\right) &= p_{13} \pi , \\
- \phi_\mathcal{T} \left(0,0,1\right) + \phi_\mathcal{T} \left(0,0,2\right) &= p_{13} \pi , \\
- \phi_\mathcal{T} \left(0,0,6\right) + \phi_\mathcal{T} \left(0,0,3\right) &= p_{13} \pi , \\
- \phi_\mathcal{T} \left(0,0,5\right) + \phi_\mathcal{T} \left(0,0,4\right) &= p_{13} \pi , \\
- p_{10} \pi - \phi_\mathcal{T} \left(0,0,4\right) + \phi_\mathcal{T} \left(0,0,5\right) &= p_{13} \pi , \\
- p_{10} \pi - \phi_\mathcal{T} \left(0,0,3\right) + \phi_\mathcal{T} \left(0,0,6\right) &= p_{13} \pi , \\
\end{align*}
\endgroup
which implies $p_{10} = 0 \, \mathrm{mod} \, 2$. We are left with five equations to determine the form of $\phi_\mathcal{T}\left(0,0,s\right)$,
\begingroup
\allowdisplaybreaks
\begin{align}
- \phi_\mathcal{T} \left(0,0,1\right) + \phi_\mathcal{T} \left(0,0,5\right) &= \left(p_2 + p_3\right) \pi , \label{bosontimereversalgaugeconstraint1} \\
- \phi_\mathcal{T} \left(0,0,2\right) + \phi_\mathcal{T} \left(0,0,4\right) &= \left(p_2 + p_3\right) \pi , \label{bosontimereversalgaugeconstraint2} \\
- \phi_\mathcal{T} \left(0,0,1\right) + \phi_\mathcal{T} \left(0,0,2\right) &= p_{13} \pi , \label{bosontimereversalgaugeconstraint3} \\
- \phi_\mathcal{T} \left(0,0,3\right) + \phi_\mathcal{T} \left(0,0,6\right) &= p_{13} \pi , \label{bosontimereversalgaugeconstraint4} \\
- \phi_\mathcal{T} \left(0,0,4\right) + \phi_\mathcal{T} \left(0,0,5\right) &= p_{13} \pi .
\label{bosontimereversalgaugeconstraint5}
\end{align}
\endgroup
Consider the sublattice dependent gauge transformation defined by
\begin{equation*}
\begin{aligned}
G_5: \: & \phi(x,y,s=1,2,4,5) = \phi_5 , \\
& \phi(x,y,s=3,6) = 0 ,
\end{aligned}
\end{equation*}
which does not affect the phase $\phi_X$ associated with space group elements $X$. It changes $\phi_\mathcal{T}$ by
\begin{equation*}
\begin{aligned}
\phi_\mathcal{T} \left(x,y,s=1,2,4,5\right) &\longrightarrow \phi_\mathcal{T} \left(x,y,s\right) + 2 \phi_5 , \\
\phi_\mathcal{T} \left(x,y,s=3,6\right) &\longrightarrow \phi_\mathcal{T} \left(x,y,s\right) .
\end{aligned}
\end{equation*}
By an appropriate choice of $\phi_5$, we can fix $\phi_\mathcal{T}\left(0,0,1\right)=0$ and $\phi_\mathcal{T}\left(0,0,5\right)=\left(p_2+p_3\right)\pi$ from \eqref{bosontimereversalgaugeconstraint1}. \eqref{bosontimereversalgaugeconstraint3} and \eqref{bosontimereversalgaugeconstraint5} then implies $\phi_\mathcal{T}\left(0,0,2\right)=p_{13} \pi$ and $\phi_\mathcal{T}\left(0,0,4\right)=\left(p_2+p_3+p_{13}\right) \pi$. Furthermore, with a similar sublattice dependent gauge transformation defined by
\begin{equation*}
\begin{aligned}
G_6: \: & \phi(x,y,s=3,6) = \phi_6 , \\ 
& \phi(x,y,s=1,2,4,5) = 0 ,
\end{aligned}
\end{equation*}
we can fix $\phi_\mathcal{T}\left(0,0,3\right)=0$ and $\phi_\mathcal{T}\left(0,0,6\right)=p_{13} \pi$ from \eqref{bosontimereversalgaugeconstraint4}.
In conclusion,
\begin{equation} \label{timereversalgaugeboson}
\phi_\mathcal{T}\left(x,y,s\right) = \phi_\mathcal{T}\left(0,0,s\right) ,
\end{equation}
with
\begin{equation*}
\begin{aligned}
\phi_\mathcal{T}\left(0,0,s=1,3\right) &= 0 , \\
\phi_\mathcal{T}\left(0,0,s=2,6\right) &= p_{13}\pi , \\
\phi_\mathcal{T}\left(0,0,s=4\right) &= \left(p_2 + p_3 + p_{13}\right)\pi , \\
\phi_\mathcal{T}\left(0,0,s=5\right) &= \left(p_2 + p_3\right)\pi . \\
\end{aligned}
\end{equation*}
Solving the bosonic PSG involving time reversal symmetry in such a manner introduce an independent $\mathbb{Z}_2$ variable $p_{13}$ on top of $p_2$ and $p_3$, which arise from consideration of spatial symmetries only. There are in total $2^3=8$ possible bosonic spin liquid states labeled by $\left(p_2,p_3,p_{13}\right)$ that respect the space group of non-symmorphic Kagome lattice and time reversal symmetry. From \eqref{timereversalgaugedefinition}, we see that $2 \, \mathrm{Arg}\lbrace A_{ij} \rbrace = -\phi_\mathcal{T}(i) - \phi_\mathcal{T}(j)$, therefore time reversal symmetry restricts an ansatz to be either real or imaginary by \eqref{timereversalgaugeboson}.

\section{\label{meanfieldcalculation} Mean Field Calculation Techniques}

The flow of a mean field iteration is outlined as follows. First, we fix the bosonic density $\kappa$. Define time step $t$ such that it is initially $0$ and increases by $1$ after an iteration is completed. At time $t=0$, we choose some random value for the mean field parameters $A_{ij}$ and $B_{ij}$. At time $t>0$, we solve for $\mu_1$ and $\mu_2$ from the constraints \eqref{numberconstraintboson1245} and \eqref{numberconstraintboson36} with the value of $A_{ij}$ and $B_{ij}$ from time $t-1$. Call the solutions $\mu_1(t)$ and $\mu_2(t)$. Then, we evaluate the expectation values $\langle \hat{A}_{ij} \rangle$ and $\langle \hat{B}_{ij} \rangle$ with the set of inputs $\lbrace \mu_1(t), \mu_2(t), A_{ij}(t-1), B_{ij}(t-1) \rbrace$, which outputs $A_{ij}(t)$ and $B_{ij}(t)$ respectively. The mean field iteration is now completed and $t$ increases by $1$. If the value of mean field parameters converge upon a sufficiently large number of iterations, then a mean field solution is obtained.

In practice, mean field theory is solved by iteration of self-consistent equations \eqref{selfconsistentequationboson} in momentum space. We perform Fourier transformation
\begin{equation} \label{fouriertransformation}
b_{i,\alpha,s} = \frac{1}{\sqrt{N}} \sum_\mathbf{k} b_{\mathbf{k},\alpha,s} e^{i \mathbf{k} \cdot \mathbf{R}_i}
\end{equation}
to obtain the mean field Hamiltonian \eqref{heisenbergmeanfieldboson}, which is written in terms of the coupling constants $J, J', J_1, J_2$ and the mean field parameters $A,B',B_1,A_2$, in momentum space. In \eqref{fouriertransformation}, $i$ labels the individual unit cells, $s=1,\ldots,6$ indexes the sublattice and $N$ is the total number of unit cells (the total number of sites is therefore $N_s=6N$). Define the $2n$-component vector spinon field \cite{PhysRevB.74.174423}
\begin{equation} \label{vectorspinonfieldboson}
\Psi_\mathbf{k} = \begin{pmatrix} \vec{b}_{\mathbf{k} \uparrow} \\ \vec{b}^\dagger_{-\mathbf{k} \downarrow} \end{pmatrix} ,
\end{equation}
where we have used the abbreviation $\vec{b}_{\mathbf{k} \uparrow} = (b_{\mathbf{k},\uparrow,1}, \ldots, b_{\mathbf{k},\uparrow,n})^\mathrm{T}$ and $\vec{b}_{-\mathbf{k} \downarrow}^\dagger = (b_{-\mathbf{k},\downarrow,1}^\dagger, \ldots, b_{-\mathbf{k},\downarrow,n}^\dagger)^\mathrm{T}$, with $n=6$ being the total number of sublattices. The mean field Hamiltonian can be expressed as
\begin{equation} \label{fourierhamiltonianboson}
\begin{aligned}[b]
H_\mathrm{MF} = & \sum_\mathbf{k} \Psi_\mathbf{k}^\dagger \mathrm{D}_\mathbf{k} \Psi_\mathbf{k} \\ 
& + 4N \left( J \left\lvert A \right\rvert^2 - J' \left\lvert B' \right\rvert^2 - J_1 \left\lvert B_1 \right\rvert^2 + J_2 \left\lvert A_2 \right\rvert^2 \right) \\
& + 4N \mu_1 \left(1+\kappa\right) + 2 N \mu_2 \left(1+\kappa\right) .
\end{aligned}
\end{equation}
$\mathrm{D}_\mathbf{k}$ is a $2n \times 2n$ matrix that depends on the coupling constants, mean field parameters and chemical potentials. Evaluating the expectation value of bond and number operator, as in \eqref{selfconsistentequationboson} and \eqref{selfconsistentequationchemical}, requires diagonalizing the Fourier transformed mean field Hamiltonian by Bogoliubov transformation, which is discussed in Appendix \ref{diagonalizeboson}.

If the parameter space is small, we can just assign arbitrary initial values to the mean field parameters and iterate the self-consistent equations until they converge. Different set of initial values should be tried to avoid identifying solution that leads to local but not global minimum of the energy. Unfortunately, since we have $8$ independent mean field parameters in this particular problem, our parameter space is considerably large (a generic complex-valued parameter counts twice because the real and imaginary part are independent). Conventional mean field iteration is no longer an ideal primary tool to tackle such problem, as it becomes difficult to check whether a solution corresponds to the global minimum, and slow convergence is expected. We need a more efficient way to explore the solution space. This is done through simulated annealing (SA), which is discussed in Appendix \ref{simulatedannealing}.

SA is basically a method of probabilistic search, which suggest an approximate solution to the global minimum. We can however refine the solution from SA by inputing it as initial condition for mean field iteration. Also, for bosonic densities $\kappa$ close to the critical value $\kappa_c$, tiny changes can easily yield numerically insensible result, which our algorithm rejects by construct, rendering the exploration of solution space difficult. Therefore, mean field iteration is still important. Our general strategy is to employ SA for several $\kappa$ sufficiently lower than $\kappa_c$, and then perform mean field iteration with solutions from SA as initial condition. We tune up $\kappa$ gradually and iterate the self-consistent equations, with the solution from previous mean field iteration as initial condition, until we reach $\kappa_c$.

\subsection{\label{diagonalizeboson} Bogoliubov Transformation}

Bogoliubov transformation allows us to diagonalize the Fourier transform matrix $\mathrm{D}_\mathbf{k}$ by introducing a new pair of annihilation and creation operators, $\gamma_{\mathbf{k}}$ and $\gamma_{\mathbf{k}}^\dagger$, which satisfy the original bosonic commutation relation \eqref{commutation}. The $b$ operators are linear combinations of them,\cite{PhysRevB.74.174423,PhysRevB.45.12377}
\begin{equation} \label{gammabogoliugovpsi}
\Psi_\mathbf{k} = \mathrm{M}_\mathbf{k} \Gamma_\mathbf{k} ,
\end{equation}
where $\Gamma_\mathbf{k}$ is the vector spinon field \eqref{vectorspinonfieldboson} of $\gamma$ operators. The $2n \times 2n$ matrix $\mathrm{M}_\mathbf{k}$ is refered to as Bogoliubov transformation matrix. Define the $2n \times 2n$ diagonal matrix in which the first $n$ nonzero entries equal to $1$ and the remaining equal to $-1$,
\begin{equation} \label{tau3}
\tau_3 = \left( \begin{array}{c|c} 1 & 0 \\ \hline 0 & -1 \end{array} \right) .
\end{equation}
From the commutation relation \eqref{commutation} among the $b$ operators, we have the following identity \cite{PhysRevB.45.12377}
\begin{equation} \label{bosoncommutation}
\left[ \Psi_\mathbf{k}^i, \Psi_{\mathbf{k}'}^{\dagger j} \right] = \delta_{\mathbf{kk}'} \tau_3^{ij} .
\end{equation}
\eqref{bosoncommutation} is also true for $\Gamma_\mathbf{k}$ since Bogoliubov transformation has to preserve the commutation relation. Then,
\begin{equation*}
\begin{aligned}
\delta_{\mathbf{kk}'} \tau_3^{ij} &= \left[ \Psi_\mathbf{k}^i, \Psi_{\mathbf{k}'}^{\dagger j} \right] \\
&= \left[ \sum_m \mathrm{M}_\mathbf{k}^{im} \Gamma_\mathbf{k}^m, \sum_n \Gamma_{\mathbf{k}'}^{\dagger n} \mathrm{M}_{\mathbf{k}'}^{\dagger nj} \right] \\
&= \sum_{mn} \mathrm{M}_\mathbf{k}^{im} \left[ \Gamma_\mathbf{k}^m, \Gamma_{\mathbf{k}'}^{\dagger n} \right] \mathrm{M}_{\mathbf{k}'}^{\dagger nj} \\
&= \sum_{mn} \delta_{\mathbf{kk}'} \mathrm{M}_\mathbf{k}^{im} \tau_3^{mn} \mathrm{M}_{\mathbf{k}'}^{\dagger nj} ,
\end{aligned}
\end{equation*}
or simply \cite{PhysRevB.45.12377}
\begin{equation} \label{bogoliubovtau}
\mathrm{M}_\mathbf{k} \tau_3 \mathrm{M}_\mathbf{k}^\dagger = \tau_3 .
\end{equation}
Diagonalization of the Fourier transform matrix $\mathrm{D}_\mathbf{k}$ requires
\begin{equation*}
\Psi_\mathbf{k}^\dagger \mathrm{D}_\mathbf{k} \Psi_\mathbf{k} = \Gamma_\mathbf{k}^\dagger \mathcal{E}_\mathbf{k} \Gamma_\mathbf{k} ,
\end{equation*}
where $\mathcal{E}_\mathbf{k} = \mathrm{diag}\,(\omega_{\mathbf{k},\uparrow,1}, \ldots, \omega_{\mathbf{k},\uparrow,n}, \omega_{-\mathbf{k},\downarrow,1}, \ldots, \omega_{-\mathbf{k},\downarrow,n})$ is the eigenvalue matrix. This implies \cite{PhysRevB.45.12377}
\begin{equation} \label{bogoliubovfourier}
\mathrm{M}_\mathbf{k}^\dagger \mathrm{D}_\mathbf{k} \mathrm{M}_\mathbf{k} = \mathcal{E}_\mathbf{k} .
\end{equation}
Combining \eqref{bogoliubovtau} and \eqref{bogoliubovfourier} gives \cite{PhysRevB.45.12377}
\begin{equation} \label{diagonalizationboson}
\mathrm{M}_\mathbf{k}^{-1} \tau_3 \mathrm{D}_\mathbf{k} \mathrm{M}_\mathbf{k} = \tau_3 \mathcal{E}_\mathbf{k} .
\end{equation}
Therefore, we are really diagonalizing $\tau_3 \mathrm{D}_\mathbf{k}$ instead of merely $\mathrm{D}_\mathbf{k}$ in the usual linear algebra sense, and the eigenvalue matrix is $\tau_3 \mathcal{E}_\mathbf{k}$. The expectation value of $H_\mathrm{MF}$ and other operators (e.g. bond operators in \eqref{selfconsistentequationboson}) is evaluated with respect to the ground state of $\gamma$ spinon defined by $\gamma_{\mathbf{k},\alpha,s} \lvert 0 \rangle = 0$ for every $\mathbf{k}$, $\alpha$ and $s$. If we express these operators in terms of $\gamma$ operators, their expectation values are given by summing certain matrix elements of $\mathrm{M}_\mathbf{k}$ over $\mathbf{k}$.

The diagonalization \eqref{diagonalizationboson} is done numerically for complicated systems like Kagome lattice in which $\mathrm{M}_\mathbf{k}$ does not admit a nice analytical expression. We recommend Refs.~\onlinecite{PhysRevB.87.125127,Copla1978} to readers who are interested in the details of computer algorithm of Bogoliubov transformation.

\subsection{\label{simulatedannealing} Simulated Annealing}

Since we have a large parameter space spanned by four independent complex-valued mean field amplitudes $A,B',B_1,A_2$, we implement simulated annealing (SA) to determine the solution that minimize the mean field energy $\langle H_\mathrm{MF} \rangle$. We follow the basic procedure outlined in Ref.~\onlinecite{Corana1987}.

We would like to minimize a multivariable cost function $f\left(x_1,\ldots,x_n\right)$ that is bounded below. First, we choose an arbitrary initial configuration $(x_1^0, \ldots, x_n^0) \in \mathbb{R}^n$ and evaluate $f_0 \equiv f(x_1^0, \ldots, x_n^0)$. Then, we explore the parameter space by perturbing the variables probabilistically, for instance $x_1^0 \longrightarrow x_1^0 + \delta$ with $\delta$ being a random perturbation. Let $\epsilon$ be a positive constant that characterizes the magnitude of $\delta$, and $p \in [0,1]$ be a random number. We require the random perturbation $\delta$ to be bounded by $\epsilon$, which can be done by choosing $\delta=(-1+2p)\epsilon$. We evaluate $f_1 \equiv f(x_1^0+\delta,x_2^0,\ldots,x_n^0)$ and compare $f_1$ with $f_0$. If $f_1 \leq f_0$, we accept the change completely. If $f_1 > f_0$, we accept the change with probability $\exp\left(-\left(f_1-f_0\right)/T\right)$, where $T>0$ is a parameter analogous to temperature in Boltzmann factor. This acceptance rule is known as Metropolis criterion. After that, we move on to perturb the remaining variables $x_2^0, \ldots, x_n^0$, one at a time, in similar fashion. If every variable is perturbed once, we say that a \textit{cycle} is completed. Carrying out a sufficiently large number of cycles while gradually decreasing the temperature, we can (hopefully) find a configuration $\left(y_1,\ldots,y_n\right)$ that well estimates the global minimum.

The temperature $T$ in the acceptance probability $\exp\left(-\left(f-f_0\right)/T\right)$ plays a crucial role in the exploration of parameter space and the convergence of solution. For large $T$ compared to the typical change in cost function, $\Delta f = \overline{f-f_0}$, fluctuations in the variables $x_i$ occur more often such that a larger portion of the solution space is explored. For small $T$, solutions that yield higher cost tend to be rejected. We usually start with large $T$ and decrease it in subsequent cycles.

In our system, the cost function is just the mean field energy,
\begin{equation} \label{energyfunction}
\langle H_\mathrm{MF} \rangle \equiv E\left(A,B',B_1,A_2\right) ,
\end{equation}
with $\kappa$ fixed at some value. The chemical potentials $\mu_1$ and $\mu_2$ are not freely varying parameters because they are solved self-consistently by \eqref{numberconstraintboson1245} and \eqref{numberconstraintboson36} for a given set of mean field amplitudes. Since each of the mean field parameters is complex-valued, we have to perturb the real and imaginary parts independently, and the solution space is effectively $\mathbb{R}^{8}$.

Convergent solution is found upon implementation of SA. As a check, we input the solution from SA as initial condition for mean field iteration, and verify that it is also the convergent solution of self-consistent equations \eqref{selfconsistentequationboson}.

\section{\label{linearapproximationscheme} Linear Approximation Scheme}
\begin{figure}[!htbp]
\includegraphics[scale=0.4]{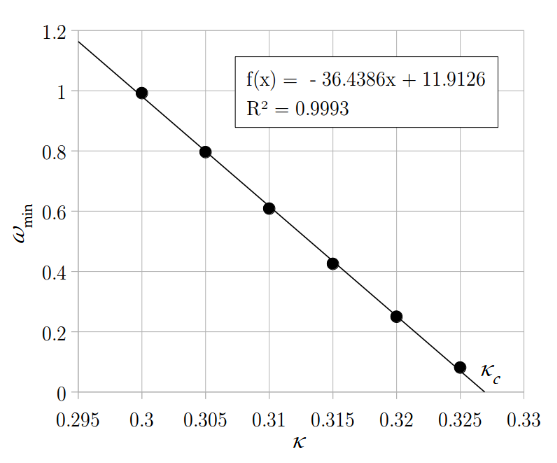}
\caption{\label{linearapproximationfigure} We determine $\kappa_\mathrm{c}$ by a linear approximation scheme. With the spin liquid state $\left(p_2,p_3\right)=\left(0,0\right)$ as example, as $\omega_\mathrm{min}$ goes to $0$ with increasing $\kappa$, we plot several data points $\left(\kappa,\omega_\mathrm{min}\right)$, add a line through them, and estimate the x-intercept as $\kappa_c$.}
\end{figure}

FIG. \ref{linearapproximationfigure} shows the linear approximation scheme we use to determine the critical bosonic density $\kappa_c$ from mean field theory. Related discussion can be found in Section \ref{bosonicspinliquid}.

\section{\label{energycomparison} Energetics of Spin Liquid States}
TABLE \ref{energycomparisontable} lists the energy of four spin liquid states $(p_2,p_3)$ that respect the space group of non-symmorphic Kagome lattice at several bosonic densities $\kappa$ below the critical value $\kappa_c$. Related discussion can be found in Section \ref{bosonicspinliquid}.
\begin{table}[!htbp]
\caption{\label{energycomparisontable} Comparison of total energy $E$ (scaled by the total number of sites $N$) of four spin liquid states $(p_2,p_3)$ at several $\kappa<\kappa_c$. The energy shown here is in unit of $10^{-2}J$, where $J$ is the strongest coupling constant. In low $\kappa$ limit, the states with same $p_2$ has exactly the same spectrum and energy. The energy begins to differ around $\kappa=0.25$. $(1,0)$ is identified to be the most energetically favorable state.}
\begin{ruledtabular}
\begin{tabular}{c|cccc}
$\kappa \backslash (p_2,p_3)$ & $(0,0)$ & $(0,1)$ & $(1,0)$ & $(1,1)$ \\ \hline
$0.05$ & -0.951587 & -0.951587 & -0.953141 & -0.953141 \\
$0.10$ & -1.973095 & -1.973095 & -1.984894 & -1.984894 \\
$0.15$ & -3.064663 & -3.064663 & -3.101634 & -3.101634 \\
$0.20$ & -4.226422 & -4.226422 & -4.307044 & -4.307044 \\
$0.25$ & -5.458552 & -5.458552 & -5.603122 & -5.603122 \\
$0.26$ & -5.713968 & -5.714021 & -5.873327 & -5.873327 \\
$0.27$ & -5.972844 & -5.973016 & -6.147242 & -6.147242 \\
$0.28$ & -6.235185 & -6.235556 & -6.425370 & -6.425271 \\
$0.29$ & -6.500999 & -6.501673 & -6.707870 & -6.707606 \\
$0.30$ & -6.770304 & -6.771429 & -6.994841 & -6.994180 \\
$0.31$ & -7.043122 & -7.044775 & -7.286310 & -7.285185 \\
$0.32$ & -7.319378 & -7.321825 & -7.582275 & -7.580489 \\
$0.33$ & $-$ 	   & -7.602579 & -7.882738 & -7.880159 \\
$0.34$ & $-$	   & -7.887169 & -8.187831 & -8.184325 \\
$0.35$ & $-$	   & -8.175661 & $-$	   & -8.492923 \\
\end{tabular}
\end{ruledtabular}
\end{table}

\bibliography{draft}
\end{document}